\documentclass{llncs}

\def\ket#1{\mathinner{|{#1}\rangle}}

\def\Ket#1{\left|#1\right>}
{\catcode`\|=\active 
  \gdef\Braket#1{\begingroup \mathcode`\|32768\let|\BraVert\left<{#1}\right>\endgroup}
}
\def\BraVert{\egroup\,\mid@vertical\,\bgroup}
{\catcode`\|=\active
  \gdef\set#1{\mathinner{\lbrace\,{\mathcode`\|"8000\let|\midvert #1}\,\rbrace}}
  \gdef\Set#1{\left\{\:{\mathcode`\|"8000\let|\SetVert #1}\:\right\}}}
\def\midvert{\egroup\mid\bgroup}
\def\SetVert{\egroup\;\mid@vertical\;\bgroup}

\usepackage{graphicx,amsfonts,amsmath,latexsym,amssymb}
\usepackage[mathcal]{euscript}
\usepackage[table]{xcolor}

\newcommand\VL[1]{#1} 	
\newcommand\VS[1]{} 	
\newcommand\VLtodo[1]{}	

\newcommand{\Z}{\mathbb{Z}}

\newcommand{\joliH}{\mathcal{H}}

\newcommand{\joliN}{\mathcal{N}}
\newcommand{\pa}[1]{\left(#1\right)}
\newcommand{\acco}[1]{\left\{#1\right\}}

\newcommand{\etal}{\emph{et al\ }{}}
\newcommand{\eg}{\emph{e.g.\ }{}}
\newcommand{\ie}{\emph{i.e.\ }{}}
\newcommand{\cf}{\emph{cf.\ }{}} 
\newcommand{\videinfra}{\emph{vide infra}{}} 

\newcommand{\Phase}{R($\frac{\pi}{4})${}}
\newcommand{\cPhase}{controlled-\Phase{}}

\newcommand{\cells}[4]{
  \centering
  \Ket{
   \,
  \begin{tabular}{ | p{2.8mm} | p{2.8mm} | }
   \hline			
     #1 & #2   \\ \hline
     #3 & #4   \\ \hline 
   \end{tabular}\,
  } 
}

\newcommand{\barrier}{\cellcolor{orange}}

\newtheorem{Th}{Theorem}

\newtheorem{Def}{Definition}

\begin{document}

\title{Intrinsically universal $n$-dimensional quantum cellular automata}

\author{Pablo Arrighi \and Jonathan Grattage}

\institute{University of Grenoble, \VL{Laboratoire }LIG,\VL{\\}
 \VL{B\^{a}timent IMAG C, }220 rue de la Chimie,\VL{\\}
 38400 \VS{SMH}\VL{Saint-Martin-d'H\`eres}, France
 \and
 Ecole Normale Sup\'erieure de Lyon, \VL{Laboratoire }LIP,\VL{\\}
 46 all\'ee d'Italie, 69364 Lyon cedex 07, France}

\maketitle

\begin{abstract}
There have been several non-axiomatic approaches taken to define Quantum Cellular Automata (QCA). Partitioned QCA (PQCA) are the most canonical of these non-axiomatic definitions. 
In this work we first show that any QCA can be put into the form of a PQCA. Our construction reconciles all the non-axiomatic definitions of QCA, showing that they can all simulate one another, and hence that they are all equivalent to the axiomatic definition.  
Next, we describe a simple $n$-dimensional QCA capable of simulating all others, in that the initial configuration and the forward evolution of any $n$-dimensional QCA can be encoded within the initial configuration of the intrinsically universal QCA, and that several steps of the intrinsically universal QCA then correspond to one step of the simulated QCA. 
Both results are made formal by defining generalised $n$-dimensional intrinsic simulation, \ie a notion of simulation which preserves the topology in the sense that each cell of the simulated QCA is encoded as a group of adjacent cells in the universal QCA. We argue that this notion brings the computer science based concepts of simulation and universality one step closer to theoretical physics. 
\end{abstract}


\section{Introduction}\label{sec:introduction}
Cellular automata (CA), first introduced by von Neumann \cite{Neumann}, consist of an array of identical cells, each of which may take one of a finite number of possible states. The whole array evolves in discrete time steps by iterating a function $G$. This global evolution $G$ is shift-invariant (it acts in the same way everywhere) and local (information cannot be transmitted faster than some fixed number of cells per time step).

\subsection{QCA: Importance and Competing Definitions} 

The modern axiomatisation of quantum theory in terms of the density matrix formalism was provided by von Neumann in 1955 \cite{NeumannQT}, who also developed the cellular automata (CA) model of computation in 1966 \cite{Neumann},  but he did not bring the two together. Feynman did so in 1986 \cite{FeynmanQC},  just as he was developing the concept of quantum computation (QC). Listed below are the key multidisciplinary motivations for studying QCA, the first two being those of Feynman. 
\VS{\\}\VL{\begin{itemize}}
\VS{-}\VL{\item} \emph{Implementation perspective.} QCA may provide an important path to realistic implementations of QC, mainly because they eliminate the need for an external, classical control over the computation and hence the principal source of decoherence. This is continuously under investigation \cite{BrennenWilliams,LloydQCA,NagajWocjan,Twamley,VollbrechtCirac}.
\VS{-}\VL{\item} \emph{Simulation perspective.} QC was first conceived as a way to efficiently simulate other quantum physical systems. Whilst other applications have been invented since, this still remains a likely and important application of QC. However, it may not be straightforward to encode the theoretical description of a quantum physical system into a QC in a relevant manner, \ie so that the QC can then provide an accurate and efficient simulation. QCA constitute a natural theoretical setting for this purpose, in particular via Quantum Lattice-Gas Automata \cite{Bialynicki-Birula,BoghosianTaylor2,Eakins,LoveBoghosian,MeyerQLGI}.
\VS{-}\VL{\item} \emph{CA perspective.} By their definition (given above), CA are shift-invariant and causal.
CA are therefore a physics-like model of computation (a term coined by Margolus \cite{MargolusPhysics}), as they share some fundamental symmetries of theoretical physics: homogeneity (invariance of physical laws in time and space), causality, and (often) reversibility. Thus it is natural, following Margolus \cite{MargolusQCA}, to study their quantum extensions.
\VS{-}\VL{\item} \emph{Models of computation perspective.} \VL{There are many models of distributed computation (\eg CCS, $\pi$-calculus), but often in such models the idea of space is not directly related to our general understanding, such as our intuitive understanding of relative positions of objects in $3$D space. These models are not adequate for reasoning about simple space-sensitive synchronisation problems, such as `machine self-reproduction' \cite{Codd,Neumann} or the `Firing Squad' problem \cite{MazoyerFiring,MooreFiring}. In contrast, }CA were initially used  to model spatially distributed computation in space \cite{ToffoliMargolusModelling}. Moreover, QCA provide a model of QC, and hence constitute a framework to model and reason about problems in spatially distributed QC.
\VS{-}\VL{\item} \emph{Theoretical physics perspective.} QCA could provide helpful toy models for theoretical physics \cite{LloydQG}. For this purpose it could build bridges between computer science and theoretical physics, as the present paper attempts to for the concept of universality.
\VL{\end{itemize}}

These motivations demonstrate the importance of studying QCA. Once this is acknowledged researchers are faced with an overabundance of competing definitions of QCA.
An examination shows that there are four main approaches to defining QCA: the axiomatic style \cite{SchumacherWerner,ArrighiLATA,ArrighiUCAUSAL}, the multilayer block representation \cite{ArrighiUCAUSAL,PerezCheung}, the two-layer block representation \cite{BrennenWilliams,NagajWocjan,SchumacherWerner,Karafyllidis,Raussendorf,VanDam}, and Partitioned QCA (PQCA) \cite{VanDam,WatrousFOCS,InokuchiMizoguchi}. A natural first questions to consider is whether they are equivalent, and in what sense.

\subsection{QCA: Simulation} \label{subsec:CA}

Probably the most well known CA is Conway's `Game of Life'; a two-dimensional CA which has been shown to be universal for computation in the sense that any Turing Machine (TM) can be encoded within its initial state and then executed by evolution of the CA \cite{winningways}. As TM are generally considered to be a robust definition of `what an algorithm is' in classical computer science, this result could be perceived as providing a conclusion to the topic of CA simulation.  However, this is not the case, as CA do more than just running any algorithm. They run distributed algorithms in a distributed manner, model phenomena together with their spatial structure, and allow the use of the spatial parallelism inherent in the model. These features, modelled by CA and not by TM, are all interesting, and so the concepts of simulation and universality needed be revisited in this context to account for space. This has been done by returning to the original meaning of the word \emph{simulation} \cite{AlbertCulik,Banks,DurandRoka}, namely the ability for one instance of a computational model to simulate other instances of the \emph{same} computational model. The introduction of a partial order on CA via  groupings \cite{MazoyerRapaport}, and subsequent generalisations \cite{OllingerJAC,Theyssier}, have led to elegant and robust definitions of intrinsic simulation. Intrinsic simulation formalises the ability of a CA to simulate another in a space-preserving manner. 
Intuitively this is exactly what is needed to show the equivalence between the various competing definitions of QCA, \ie that they can all simulate each other in a space-preserving manner. The definition of intrinsic simulation has already been translated in the quantum context \cite{ArrighiFI}, however as it stands this is not sufficient to obtain the desired result. In this paper the definition of intrinsic simulation in the quantum context is discussed and developed, before the equivalence between all the various above-mentioned definitions of QCA is tackled. 

\subsection{QCA: Simplification} \label{subsec:literature}

Intrinsic universality is the ability to intrinsically simulate any other QCA. Here we show that the axiomatic style QCA, the multilayer block representation QCA, the two-layer block representation QCA, and the PQCA are equivalent, entailing that PQCA are intrinsically universal. Here the PQCA is chosen as the prime model as it is
the simplest way to describe a QCA. Therefore, the result developed in this work is also a simplifying one for the field of QCA as a whole. From a theoretical physics perspective, showing that `Partitioned Quantum Cellular Automata are universal' is a statement that `scattering phenomena are universal physical phenomena'.

There are several related results in the CA literature. Several influential works by Morita \etal  emphasise Reversible Partitioned CA universality. For instance, they  provide computation universal Reversible Partitioned CA constructions \cite{MoritaCompUniv1D,MoritaCompUniv2D}, and their ability to simulate any CA in the one-dimensional case is also shown \cite{MoritaIntrinsicUniv1D}. The problem of intrinsically universal Reversible CA (RCA) constructions was tackled by Durand-Lose \cite{Durand-LoseLATIN,Durand-LoseIntrinsic1D}. The difficulty is in having an $n$-dimensional RCA simulate all other $n$-dimensional RCA and not, say, the $(n-1)$-dimensional RCA, otherwise a history-keeping dimension could be used, as in Toffoli \cite{ToffoliConstruction}. Strongly related to this is the work on block representations of RCA by Kari \cite{KariCircuit}.

\subsection{QCA: Universality}

The QCA-related results are focused on universality. Watrous \cite{WatrousFOCS} proved that QCA are universal in the sense of QTM. Shepherd, Franz and Werner \cite{ShepherdFranz} defined a class of QCA where the scattering unitary $U_i$ changes at each step $i$ (classical control QCA). Universality in the circuit-sense has already been achieved by Van Dam \cite{VanDam}, Cirac and Vollbrecht \cite{VollbrechtCirac}, Nagaj and Wocjan \cite{NagajWocjan}, and Raussendorf \cite{Raussendorf}. In the bounded-size configurations case, circuit universality coincides with intrinsic universality, as noted by Van Dam \cite{VanDam}. Intrinsically universal QCA in the one-dimensional case have also been resolved \cite{ArrighiFI}. Given the crucial role of this in classical CA theory \cite{Durand-LoseEnc}, the issue of intrinsic universality in the $n$-dimensional case needed to be addressed. Having then shown that PQCA, a simple subclass of QCA, are intrinsically universal, it remained to show that there existed a $n$-dimensional PQCA capable of simulating all other $n$-dimensional PQCA for $n>1$, which we show in this paper.

PQCA are QCA of a particular form, where incoming information is scattered by a fixed unitary $U$ before being redistributed, and this paper shows PQCA to be intrinsically universal. Hence the problem of finding an intrinsically universal PQCA reduces to finding some universal scattering unitary $U$ (as made formal in section \ref{subsec:flat}, Fig.~\ref{fig:flattening34}). Also, the requirements on $U$ are much more stringent than just quantum circuit universality, as the simulation of a QCA $H$ has to be done in a parallel, space-preserving manner. Moreover,  not only a single iteration of $H$ has to be simulated, but several ($H^2$, \ldots), so that after every simulation the universal PQCA is ready for a further iteration.
From a computer architecture point of view, this problem can be recast in terms of finding some fundamental quantum processing unit which is capable of simulating any other network of quantum processing units, in a space-preserving manner. From a theoretical physics perspective, this amounts to specifying a scattering phenomenon that is capable of simulating any other, again in a space-preserving manner. 
These difficulties can be overcome. A key result shown here is the construction of a simple intrinsically universal $n$-dimensional QCA.

\subsection{Layout}

The necessary theoretical background for understanding QCA, and  hence the problems addressed by this paper, is provided in section \ref{definitions}.  Intrinsic simulation is discussed and generalised in section \ref{subsecsim}. In section \ref{sec:struc} the various alternative definitions of QCA are shown to be equivalent to the simplest definition, \ie PQCA. In section \ref{sec:nuqca} a simple example of an intrinsically universal PQCA is developed. Section \ref{sec:discussion} concludes with a discussion and ideas for future directions. This paper also integrates the contributions of two already-published conference papers \cite{ArrighiPQCA,ArrighiSimple}.

\section{Definitions} \label{definitions}
\subsection{$n$-Dimensional QCA}\label{subsecdef}

This section provides the axiomatic style definitions for $n$-dimensional QCA. 
Configurations hold the basic states of an  entire array of cells, and hence denote the possible basic states of the entire QCA: 
\begin{Def}[Finite configurations]
A \emph{(finite) configuration} $c$ over $\Sigma$ is a function $c: \Z^n \longrightarrow \Sigma$, with 
$(i_1,\ldots,i_n)\longmapsto c(i_1,\ldots,i_n)=c_{i_1\ldots i_n}$, such that there exists a (possibly empty)
finite set $I$ satisfying $(i_1,\ldots,i_n)\notin I\Rightarrow c_{i_1\ldots i_n}=q$, where $q$ is a distinguished \emph{quiescent} state of $\Sigma$.
The set of all finite configurations over $\Sigma$ will be denoted $\mathcal{C}^{\Sigma}_{fin}$.
\end{Def}

Since this work relates to QCA rather than CA, the global state of a QCA can be a superposition of these configurations. 
To construct the separable Hilbert space of superpositions of configurations the set of configurations must be countable. 
Thus finite, unbounded, configurations are considered. The quiescent state of a CA is analogous to the blank symbol of a Turing machine  tape. 

\begin{Def}[Superpositions of configurations]\label{superp} 
Let $\mathcal{H}_{\mathcal{C}^{\Sigma}_{fin}}$ be the Hilbert space of configurations. Each finite configuration $c$ is associated with a unit vector $\ket{c}$, such that the family $\pa{\ket{c}}_{c\in\mathcal{C}^{\Sigma}_{fin}}$ is an orthonormal basis of $\mathcal{H}_{\mathcal{C}^{\Sigma}_{fin}}$. A \emph{superposition of configurations} is then a unit vector in $\mathcal{H}_{\mathcal{C}^{\Sigma}_{fin}}$. 
\end{Def}

\begin{Def}[Unitarity]\label{unitarity} A linear operator $G:\mathcal{H}_{\mathcal{C}^{\Sigma}_{fin}}\longrightarrow\mathcal{H}_{\mathcal{C}^{\Sigma}_{fin}}$ is \emph{unitary} if and only if $\{G\ket{c}\,|\,c\in\mathcal{C}^{\Sigma}_{fin}\}$ is an orthonormal basis of $\mathcal{H}_{\mathcal{C}^{\Sigma}_{fin}}.$
\end{Def}

\begin{Def}[Shift-invariance]\label{shift-invariance} 
Consider the shift operation, for $k\in$\\
$\acco{1,\ldots, n}$, which takes configuration $c$ to $c'$ where for all $(i_1,\ldots ,i_n)$ we have $c'_{i_1\ldots i_k \ldots i_n}=c_{i_1\ldots i_k+1 \ldots i_n}$. Let $\sigma_k:\mathcal{H}_{\mathcal{C}^{\Sigma}_{fin}}\longrightarrow\mathcal{H}_{\mathcal{C}^{\Sigma}_{fin}}$ denote its linear extension to a superpositions of configurations. A linear operator $G:\mathcal{H}_{\mathcal{C}^{\Sigma}_{fin}}\longrightarrow\mathcal{H}_{\mathcal{C}^{\Sigma}_{fin}}$ is said to be 
\emph{shift invariant} if and only if $G\sigma_k=\sigma_k G$ for each $k$.
\end{Def}

The following definition captures the  causality of the dynamics. Imposing the condition that the state associated to a cell
(its reduced density matrix) is a function of the neighbouring cells is equivalent to stating that information
propagates at a bounded speed. 
\begin{Def}[Causality]\label{locality} 
A linear operator $G:\mathcal{H}_{\mathcal{C}^{\Sigma}_{fin}}\longrightarrow\mathcal{H}_{\mathcal{C}^{\Sigma}_{fin}}$ is said to be 
 \emph{causal}{} 
if and only if for any
$(i_1,\ldots,i_n)\in\Z_n$,  there exists a function $f$ such that $\rho'|_\joliN = f (\rho|_\joliN)$
for all $\rho$ over $\mathcal{H}_{\mathcal{C}^{\Sigma}_{fin}}$, where:\\
$\joliN=\{i_1,i_1+1\}\times\ldots\times\{i_n,i_n+1\}$, $\rho|_\joliN$ means the restriction of $\rho$ to the neighbourhood $\joliN$ in the sense of the partial trace, and $\rho' = G \rho G^\dagger$.
\end{Def}

In the classical case, the definition is that the letter to be read in some given cell $i$ at time $t+1$ depends only on the state of the cells $i$ to $i+1$ at time $t$. Transposed to a quantum setting, the above definition is obtained. To know the state of cell number $i$, only the states of cells $i$ and $i+1$ before the evolution need be known.

More precisely, this restrictive definition of causality is known in the classical case as a $\frac{1}{2}$-neighbourhood cellular automaton,  because the most natural way to represent such an automaton is to shift the cells by $\frac{1}{2}$ at each step, so that visually the state of a cell depends on the state of the two cells under it. This definition of causality is not restrictive, as by grouping cells into ``supercells'' any CA with an arbitrary finite neighbourhood $\joliN$ can be made into a $\frac{1}{2}$-neighbourhood CA. The same method can be applied to QCA, so this definition of causality holds without loss of generality.
However, the $f$ in the above definition does not directly lead to a constructive definition of a cellular automaton, unlike the local 
transition function in the classical case \cite{ArrighiUCAUSAL}.

This approach leads to the following definition of an $n$-dimensional QCA. It has been given previously \cite{ArrighiLATA,ArrighiUCAUSAL}, but clearly stems from an equivalent definition in the literature, phrased in terms of homomorphism of a $C^*$-algebra \cite{SchumacherWerner}.
\begin{Def}[QCA]\label{def:qca} 
An $n$-dimensional quantum cellular automaton (QCA) is an operator $G:\mathcal{H}_{\mathcal{C}^{\Sigma}_{fin}}\longrightarrow\mathcal{H}_{\mathcal{C}^{\Sigma}_{fin}}$
which is unitary, shift-invariant and causal.
\end{Def}

Whilst this is clearly the natural, axiomatic definition QCA, it remains a non-constructive one. In this sense it can be compared to the Curtis-Hedlund \cite{Hedlund} definition of CA as the set of continuous, shift-invariant functions. These definitions characterise (Q)CA via the global, composable properties that they must have; but they do not provide an operational, hands-on description of their dynamics.

\subsection{Multilayer Block Representation}
What is meant by an operational description of a QCA? A central tool and concept in this paper is that of a (multilayer) block representation of QCA. Intuitively, we say that a QCA $G$ admits a block representation when it can be expressed as blocks, \ie local unitaries, composed in space (via the tensor product) and time (via operator composition), thereby forming a finite-depth quantum circuit infinitely repeating across space. The structure theorem given in previous work \cite{ArrighiUCAUSAL} states that any QCA can in fact be represented in such a way:

\begin{Th}[$n$-dimensional QCA multilayer block representation]\label{th:multilayers}~\\
Let $G$ be an $n$-dimensional QCA with alphabet $\Sigma$. Let $E$ be an isometry from $\joliH_\Sigma\to\joliH_\Sigma\otimes\joliH_\Sigma$ such that $E\ket{\psi_x}=\ket{q}\otimes\ket{\psi_x}$. This mapping can be trivially extended to whole configurations, yielding a mapping $E:\joliH_{C^{\Sigma}_{fin}}\to\joliH_{C^{\Sigma^2}_{fin}}$. There then exists an $n$-dimensional QCA $H$ on alphabet $\Sigma^2$, such that $HE=EG$, and $H$ admits an $2^n$-layer block representation. Moreover $H$ is of the form 
\begin{align}
H=(\bigotimes S)(\prod K_x) \label{eq:luqca}
\end{align}
where:
\begin{itemize}
\item $(K_x)$ is a collection of commuting unitary operators all identical up to shift, each localised upon each neighbourhood $\joliN_x$;
\item $S$ is the swap gate over $\joliH_\Sigma\otimes\joliH_\Sigma$, hence localised upon each node $x$.
\end{itemize}
\end{Th}

This theorem therefore bridges the gap between the axiomatic style definition of QCA and the operational descriptions of QCA. Again, it should be compared with the Curtis-Hedlund \cite{Hedlund} theorem, which shows the equivalence between the axiomatic definition of CA and the more operational, standard definition, with a local function applied synchronously across space. One can argue that the form given in Eqn.~\ref{eq:luqca} is not that simple. A contribution of this paper is to simplify it down to PQCA.

Amongst the operational definitions of QCA listed in section \ref{sec:introduction},
only that of Perez-Delgado and Cheung \cite{PerezCheung} is not two-layer. 
They directly state, after some interesting informal arguments, that QCA are of a form similar
to that given in Eqn.~\ref{eq:luqca}.

In other words, this theorem demonstrates that starting from an axiomatic definition of QCA, such as Shumacher and Werner's \cite{SchumacherWerner}, one can derive a
circuit-like structure for $n$-dimensional QCA, thereby extending their result to $n$ dimensions. It also
demonstrates that operational definitions \cite{PerezCheung} can be given a rigorous axiomatics. 
These factors demonstrate that the definitions of P\'erez-Delgado and Cheung \cite{PerezCheung} and Shumacher and Werner \cite{SchumacherWerner} are actually equivalent, up to ancillary cells.

This shows that the axiomatic definition of QCA given in section \ref{subsecdef} is equivalent to a multilayer block representation.
There are, however, several other definitions of QCA, \ie two-layer block representations and PQCA. 
The aim is to now show that all definitions of QCA  can be reconciled via
intrinsic simulation. A quantum version of intrinsic simulation has already been developed \cite{ArrighiFI},
but only for one-dimensional QCA, and it is not general enough to state the required equivalence. 
This difficulty is addressed in the next section, where a new concept of intrinsic simulation for $n$-dimensional QCA
is developed with the required properties.

\section{Intrinsic Simulation of $n$-Dimensional QCA} \label{subsecsim}

Intrinsic simulation of one CA by another was discussed informally in section \ref{subsec:CA}. A
pedagogical discussion in the classical case was given by Ollinger \cite{OllingerJAC}, and
quantised intrinsic simulation has been formalised in the one-dimensional case \cite{ArrighiFI}. 
This definition is extended to $n$-dimensions (and relaxed, see details below) here.
The potential use of this concept in theoretical physics is also discussed.

Intuitively, `$G$ simulates $H$' is shown by translating the contents of each cell of $H$ into cells of $G$,
running $G$, and then reversing the translation;  this three step process amounts to running $H$. 
This translation should be simple (it should not provide a ``hidden'' way to compute $G$),  should preserve the topology (each cell of $H$ is encoded into cells of $G$ in a way which preserves neighbours), and  should be faithful (no information should be lost in translation). This latter requirement relates to the \emph{isometry} property of quantum theory, \ie an inner product preserving evolution with $Enc^\dagger Enc=\mathbb{I}$. This same requirement agrees with the translation being a physical process. The following definitions are thus derived.

\begin{Def}[Isometric coding]\label{isomcode} 
Consider $\Sigma_G$ and $\Sigma_H$, two alphabets with distinguished quiescent states $q_G$ and $q_H$, and such that $|\Sigma_H|\leq|\Sigma_G|$. Consider $\mathcal{H}_{\Sigma_G}$ and $\mathcal{H}_{\Sigma_H}$ the Hilbert spaces having these alphabets as their basis, and $\mathcal{H}_{\mathcal{C}_{fin}^{G}}$, $\mathcal{H}_{\mathcal{C}_{fin}^{H}}$ the Hilbert spaces of finite configurations over these alphabets.\\
Let $E$ be an isometric linear map from $\mathcal{H}_{\Sigma_H}$ to $\mathcal{H}_{\Sigma_G}$ which preserves quiescence, \ie such that $E\ket{q_H}=\ket{q_G}$. It trivially extends into an isometric linear map $Enc=(\bigotimes_{\mathbb{Z}^n} E)$ from $\mathcal{H}_{\mathcal{C}_{fin}^{H}}$ into $\mathcal{H}_{\mathcal{C}_{fin}^{G}}$, which we call an isometric encoding.\\
Let $D$ be an isometric linear map from $\mathcal{H}_{\Sigma_G}$ to $\mathcal{H}_{\Sigma_H}\otimes\mathcal{H}_{\Sigma_G}$ which also preserves quiescence, in the sense that $D\ket{q_G}=\ket{q_H}\otimes\ket{q_G}$. It trivially extends into an isometric linear map $Dec=(\bigotimes_{\mathbb{Z}^n} D)$ from $\mathcal{H}_{\mathcal{C}_{fin}^{G}}$ into $\mathcal{H}_{\mathcal{C}_{fin}^{H}}\otimes\mathcal{H}_{\mathcal{C}_{fin}^{G}}$, which we call an isometric decoding.\\
The isometries $E$ and $D$ define an isometric coding if the following condition is satisfied:\\
$\forall \ket{\psi}\in \mathcal{H}_{\mathcal{C}_{fin}^{H}},\,\exists \ket{\phi}\in \mathcal{H}_{\mathcal{C}_{fin}^{G}}\quad/\quad\ket{\psi}\otimes\ket{\phi}=Dec\pa{Enc \ket{\psi}}.$
\end{Def}

(Here $Dec$ is understood to morally be an inverse function of $Enc$, but some garbage $\ket{\phi}$ may be omitted.)

\begin{Def}[Direct simulation]\label{directsim}
Consider $\Sigma_G$ and $\Sigma_H$, two alphabets with distinguished quiescent states $q_G$ and $q_H$, and two QCA $G$ and $H$ over these alphabets. We say that $G$ directly simulates $H$, if and only if there exists an isometric coding such that\\
$\forall i\in\mathbb{N},\,\forall \ket{\psi}\in \mathcal{H}_{\mathcal{C}_{fin}^{H}},\,\exists \ket{\phi}\in \mathcal{H}_{\mathcal{C}_{fin}^{G}}\quad/\quad (G^i\ket{\psi})\otimes\ket{\phi}=Dec \pa{{H}^i\pa{Enc \ket{\psi}}}.$
\end{Def}

\noindent Unfortunately this is not enough for intrinsic simulation, as it implies that $|\Sigma_H| = |\Sigma_G|$. It is often desirable  that $G$ simulates $H$ even though the translation:\\
- takes several cells of $H$ into several cells of $G$;\\
- demands several steps of $G$ in order to simulate several steps of $H$.\\
Hence the grouping of cells is required.
\begin{Def}[Grouping]\label{def:packmap} 
Let $G$ be an $n$-dimensional QCA over alphabet $\Sigma$. Let $s$ and $t$ be two integers, $q'$ a word in $\Sigma'=\Sigma^{s^n}$. Consider the iterate global evolution $G^t$ up to a grouping of each hypercube of $s^n$ adjacent cells into one supercell. If this operator can be considered to be a QCA $G'$ over $\Sigma'$ with quiescent symbol $q'$, then we say that $G'$ is an $(s,t,q')$-grouping of $G$.
\end{Def}

A natural way to continue would be to define an intrinsically universal QCA. However, due to the continuity of $\mathcal{H}$, this approximation
can only be up to $\epsilon$. In section \ref{sec:nuqca} we provide a universal QCA with a bound on the finite error.
\begin{Def}[Intrinsic simulation]\label{def:intsim}
Consider $\Sigma_G$ and $\Sigma_H$, two alphabets with distinguished quiescent states $q_G$ and $q_H$, and two QCA $G$ and $H$ over these alphabets. We say that $G$ intrinsically simulates $H$ if and only if there exists $G'$, some grouping of $G$, and $H'$, some grouping of $H$, such that $G'$ directly simulates $H'$.
\end{Def}

In other words, $G$ intrinsically simulates $H$ if and only if there exists some isometry $E$ which translates supercells of $H$ into supercells of $G$, such that if $G$ is iterated and then translated back, the whole process is equivalent to an iteration of $H$. {This understanding is shown schematically in Fig.~\ref{IntrinsicSim}.
\begin{figure}
\centering
\includegraphics[scale=.9, clip=true, trim=0cm 0cm 0cm 0cm]{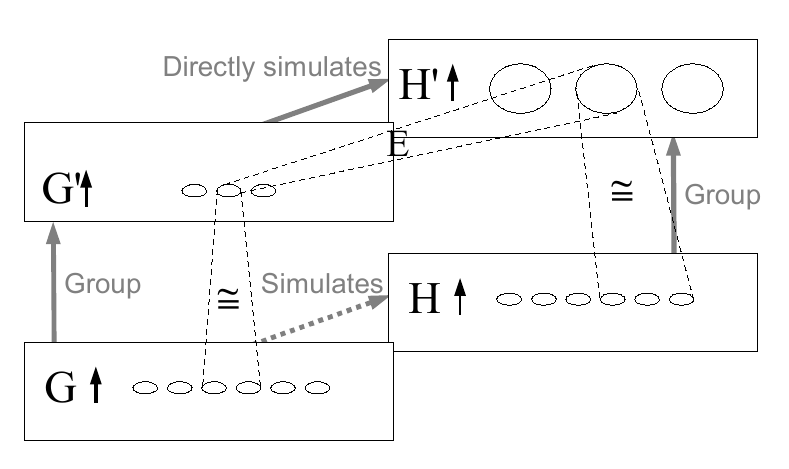}
\caption{The concept of intrinsic simulation made formal.\label{IntrinsicSim}}
\end{figure}
}

Compared with previous work \cite{ArrighiFI}, the concept of intrinsic simulation has been modified to allow the grouping
in Fig.~\ref{IntrinsicSim} on the simulated QCA side, and this variation is important to Thm.~\ref{th:pqca}.
This is analogous to the classical case \cite{Theyssier}. 

A natural way to follow would be to define the notion of an intrinsically universal QCA. However due to the continuous nature of the underlying Hilbert spaces, no QCA can be intrinsically universal in an exact sense. We can only hope to have a `dense' QCA, \ie one which can simulate any other up to some precision $\epsilon$, which can then be made arbitrarily small. In Section \ref{sec:nuqca} we provide such a construction, together with bounds on $\epsilon$.

The study of QC aims to address the issues related to the physical nature of computing, and 
over the last twenty years there have been a number of quantisations of the classical models of computation, 
and novel results on the complexity of the tasks that can be encoded in these models. It could be said that 
theoretical physics has aided theoretical computer science via this path.
Within this context, it is likely that the reverse path could also be productive. This would be part of a bigger trend where theoretical physics departs from looking at `matter' (particles interacting, scattering, forces, etc.) and seeks to look at `information' (entropy, observation, information exchanges between systems, etc.), in an attempt to clarify its own concepts. An example of this is the huge impact that quantum information theory has had on the understanding of foundational concepts such as entanglement \cite{DurVidalCirac} and decoherence \cite{PazZurek}. A computer science based approach can help to understand physical principles, not only in terms of `information', but also in terms of the `dynamics of information', \ie information processing. 

Looking at computer science, a fundamental concept in computation theory is universality. An instance of a model of computation is universal if it can simulate any other; this would also be a useful concept in physics. For example, if trying to reconcile two rather different mechanics (quantum theory and general relativity, say), finding such a \emph{minimal, universal physical phenomenon} would provide something simple to frame, so that the focus can be on reconciling the mechanics, while rich enough to guarantee that some arbitrarily complex phenomenon can be incorporated into this reconciled mechanics.  

However, the following must be considered:
\begin{itemize}
\item Firstly, a universal TM should be able to simulate each object independently in its own space. The universal physical phenomenon should be some elementary unit of computation that can be 
combined to form a 3D network, accounting for space and interactions across space satisfactorily.
\item Secondly, the universal TM is slow at simulating quantum physical phenomena, which suggests that it is not rich enough.
The universal physical phenomenon should therefore be a universal model of \emph{quantum} computation, which accounts for the \emph{cost} of simulation.
\end{itemize}
The work that has been presented in this section formalises an idea of universality which fits both these criteria, namely intrinsic universality over QCA.  
 
\VL{
\section{Constructions} \label{sec:struc}
Now that an appropriate notion of intrinsic simulation has been developed, the problem of showing an equivalence between the
different operational definitions of QCA is addressed here.

\subsection{Down to two layers: Block QCA}\label{subsec:twolayers}
Quantisations of block representations of CA are generally presented as two-layer; \cf \cite{BrennenWilliams,Karafyllidis,NagajWocjan,Raussendorf,SchumacherWerner,VanDam}. This is captured by the definition of a Block QCA (BQCA), where $\mathcal{H}^{\otimes 2^n}$ is $\mathcal{H}\otimes\ldots\otimes\mathcal{H}$, repeated $2^n$ times:
\begin{Def}[BQCA]\label{def:bqca}
A block $n$-dimensional quantum cellular automaton (BQCA) is defined by two unitary operators $U_0$ and $U_1$ such that $U_i:\mathcal{H}_{\Sigma}^{\otimes 2^n}\longrightarrow\mathcal{H}_{\Sigma}^{\otimes 2^n}$, and $U_i\ket{qq\ldots qq}=\ket{qq\ldots qq}$, \ie each takes $2^n$ cells into $2^n$ cells and preserves quiescence. Consider $G_i=(\bigotimes_{2\mathbb{Z}^n} U_i)$ the operator over $\mathcal{H}$. The induced global evolution is $G_0$ at odd time steps, and $\sigma G_1$ at even time steps, where $\sigma$ is a translation by one in all directions (Fig.~\ref{fig:structureBQCA}). 
\end{Def}

\begin{figure}
\centering
\includegraphics[scale=1.1, clip=true, trim=0cm 0cm 0cm 0cm]{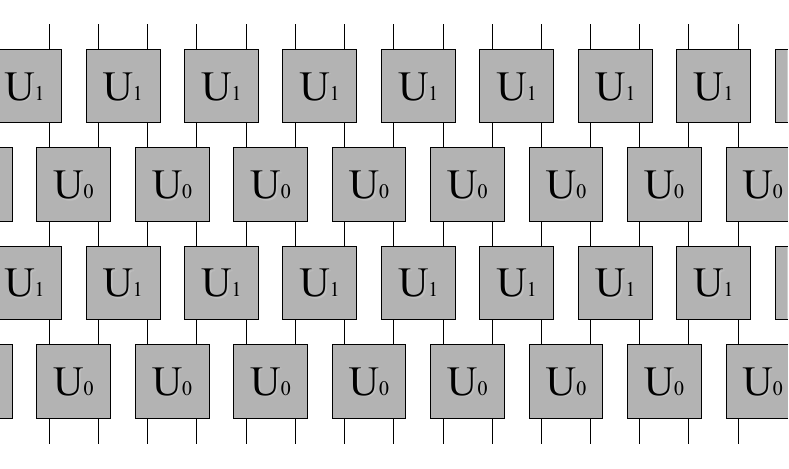}
\caption{BQCA.\label{fig:structureBQCA} The elementary unitary evolutions $U_0$ and $U_1$ are alternated repeatedly as shown, in 1D.}
\end{figure}

Showing the equivalence of the QCA and BQCA axiomatics is not trivial. 
In one direction this is simple, as 
BQCA are unitary, causal, and shift-invariant, and hence fall under the axiomatics 
and Thm.~\ref{th:multilayers}  (strictly speaking we need to group each hypercube of $2^n$ adjacent cells into a supercell,
see Def.~\ref{def:packmap}.) 
However, there are several factors to consider regarding the ability of BQCA to simulate any QCA, which are now addressed. 

In the form given by Thm.~\ref{th:multilayers}, each cell $x$ at time $t$ is successively involved in $2^n$ computations governed by a local unitary $K$, whose aim is to compute the next state of a cell within a radius $\frac{1}{2}$ from $x$ at time $t+1$. In two dimensions, a cell $x$ uses the cells West, North-West and North to work out its North-West successor, and then the cells North, North-East, East of it to compute the North-East successor (Similarly for the South-East and the South-West successors). To mimic this with a BQCA,  each original cell can be encoded into four cells, arranged so that the original cell $x$ starts in the North-West quadrant of the four cells. The first layer of the BQCA  applies the local unitary $K$ to compute the North-West successor of $x$. The second layer of the BQCA moves the original cell $x$ in the North-West quadrant. Each full application of the evolution of the BQCA corresponds only to one layer $(\bigotimes K)$, hence it will take four steps for this BQCA to simulate one step of the QCA. Fig.~\ref{fig:2layersSketch} shows a sketch of the method used. 
\begin{figure}[h!]
\centering
\includegraphics[scale=.75, clip=true, trim=0cm 0cm 0cm 0cm]{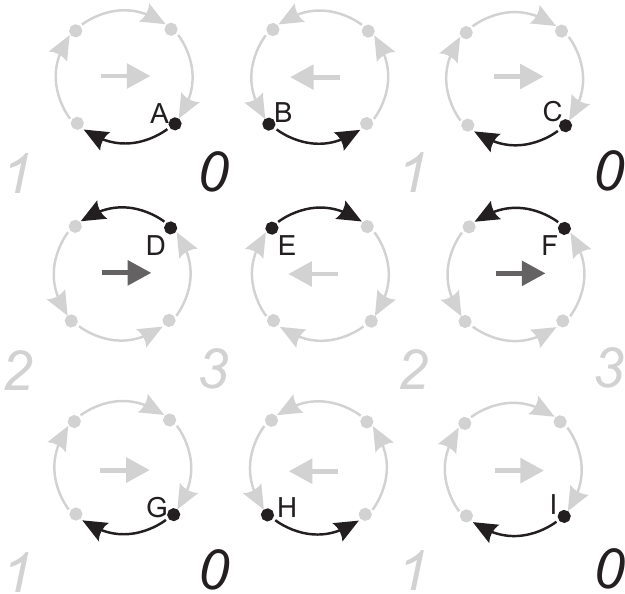}
\caption{Sketch of a BQCA simulating a QCA.\label{fig:2layersSketch} The original cell $x$ is coded into four cells, at the centre ($E$). It starts by considering the North-West as at time $0$ it will compute its North-West successor, and then move clockwise. At time $1$ it will compute its North-East successor, etc.}
\end{figure}

There are some considerations to be discussed. When cell $x$ is turning clockwise in the example, the cell to its North is turning anticlockwise. Hence we need some ancillary data coding for the path to be taken by the original cell $x$ within the four coding cells. 
Also, Thm.~\ref{th:multilayers} finishes with a $Swap$ between the `computed tape', where the results have been stored, and the `uncomputed tape', (\ie what remains of the original cell after having computed all of its successors) which is not shown in the sketch.
Hence the number of layers of $K$ computed so far has to be tracked, so that the $Swap$ occurs at the appropriate step. The $Swap$ also needs to know where the results have been stored in order to move them correctly. All of this has to be arranged spatially and efficiently, and one such method is shown in Figs. \ref{fig:2layersBig} and \ref{fig:2layersOps}. 
\begin{figure}[h!]
\centering
\VS{\includegraphics[scale=.5, clip=true, trim=0cm 0cm 0cm 0cm]{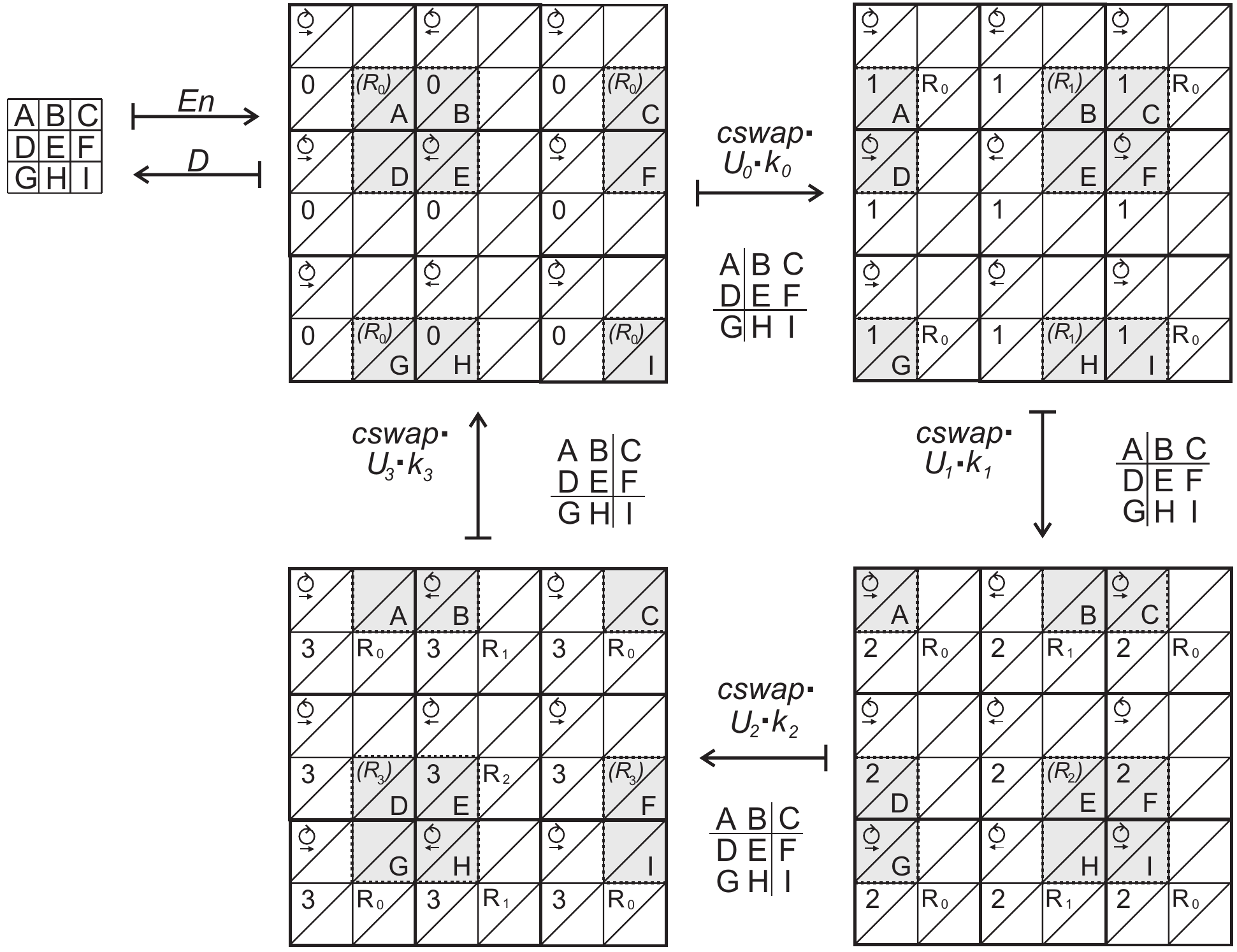}}
{\includegraphics[scale=.6, clip=true, trim=0cm 0cm 0cm 0cm]{img/QCA2.pdf}}
\caption{BQCA simulating a QCA. The grey areas denote the neighbourhood where the action of $k_x$, the first layer of the BQCA, will be significant -- \ie a group of four cells where it will perform a $K_x$ operation to work out a successor. Where this successor will be stored is indicated by $(R_x)$. At the next step $R_x$ has appeared, and the registers have been reshuffled due to the second layer of the BQCA, which acts according to the rotation-direction mark. The second layer also increases the clock count and includes the final swapping step, which only happens at time $3$. There it  ensures that $R_0$ becomes $A$, $R_1$ becomes $B$, etc. Which registers are to be swapped with one another can be calculated from the rotation and arrow marks. Each step is made formal by Fig.~\ref{fig:2layersOps}.\label{fig:2layersBig}}
\end{figure}
\begin{figure}[h!]
\centering
\includegraphics[scale=.7, clip=true, trim=0cm 0cm 0cm 0cm]{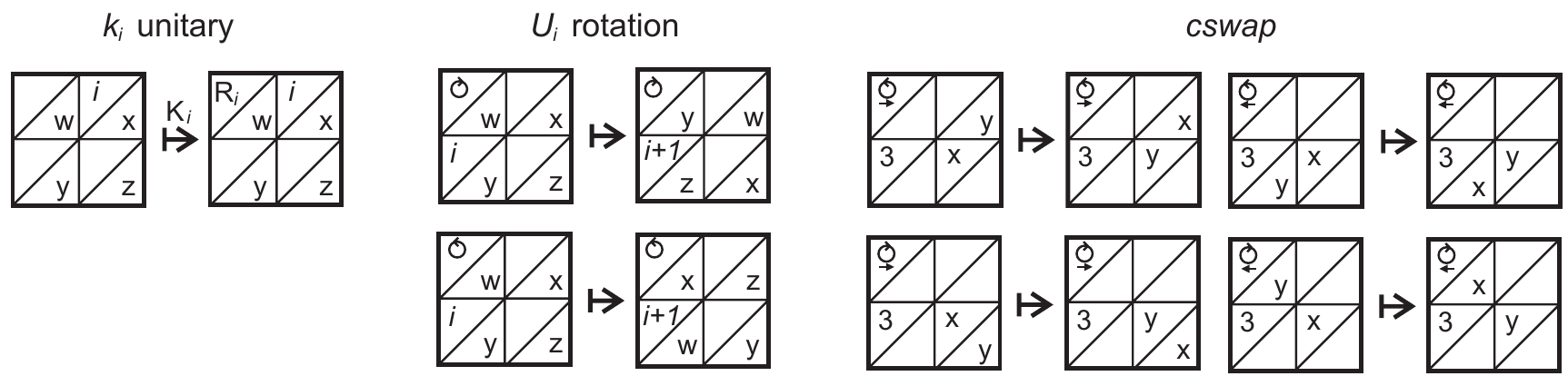}
\caption{Operations used in Fig.~\ref{fig:2layersBig}.  $k$ applies a $K$ operation whenever some data is present (data carries an extra bit to distinguish it from $\ket{q}$, say). The $U$ operation reshuffles the data by rotating it in the direction given by the indicator in the top left (clockwise or anticlockwise), and increments the index counter. Finally, $cswap$ acts as the identity in all cases except when the index is 3, when it swaps the result of the computations with the data, ready for the next round.\label{fig:2layersOps}}
\end{figure}

BQCA can therefore simulate QCA up to a relatively simple encoding, using blocks of
four cells. This explains the need for grouping on the simulated QCA side in the revised quantised intrinsic simulation, as in Fig.~\ref{IntrinsicSim}.
Encoding groups of cells rather than individual cells is also required for the PQCA discussion (\videinfra). 
This encoding is given for two dimensions, but the construct clearly generalises to $n$-dimensions. Hence QCA (Def.~\ref{def:qca})
provide a rigorous axiomatics for BQCA (Def.~\ref{def:bqca}), and BQCA provide a convenient operational description of QCA. We have shown that:
\begin{Th}[BQCA are universal]\label{th:bqca}
Given any $n$-dimensional QCA $H$, there exists an $n$-dimensional BQCA $G$ which simulates $H$.
\end{Th}

\subsection{Down to One Scattering Unitary: PQCA}\label{subsec:onescattering}

\begin{figure}[h]
\centering
\includegraphics[scale=.9, clip=true, trim=0cm 0cm 0cm 0cm]{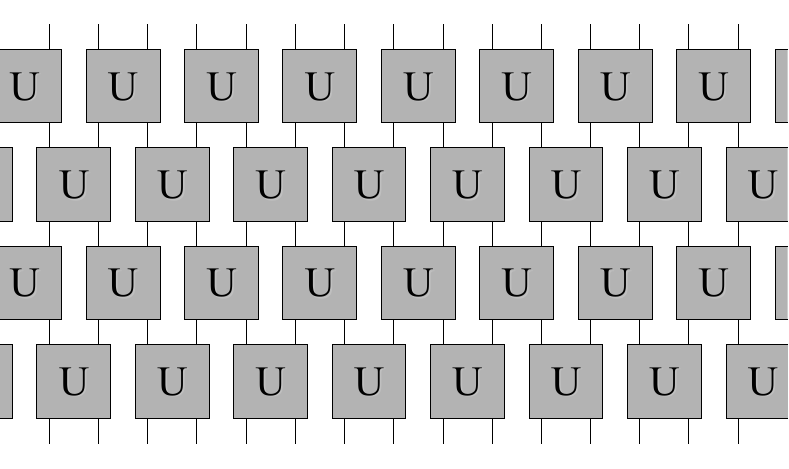}
\caption{\label{fig:structure} Partitioned one-dimensional QCA with scattering unitary $U$. Each line represents a quantum system, in this case a whole cell. Each square represents a scattering unitary $U$ which is applied to two cells. Time flows upwards.}
\end{figure}

Quantisations of partitioned representations of CA are given in several works \cite{InokuchiMizoguchi,VanDam,WatrousFOCS}. 
These constitute the simplest approach to defining QCA.
It is therefore interesting to consider whether QCA (as in Def.~\ref{def:qca}) provide a rigorous axiomatics for PQCA, and if PQCA provide a convenient operational description of QCA. A PQCA is essentially a BQCA where the two layers apply the same unitary operation, shifted appropriately.
\begin{Def}[PQCA]\label{def:pqca}
A partitioned  $n$-dimensional quantum cellular automaton (PQCA) is defined by a scattering unitary operator $U$ such that $U:\mathcal{H}_{\Sigma}^{\otimes 2^n}\longrightarrow\mathcal{H}_{\Sigma}^{\otimes 2^n}$, and $U\ket{qq\ldots qq}=\ket{qq\ldots qq}$, \ie that takes
a hypercube of $2^n$ cells into a hypercube of $2^n$ cells and preserve quiescence. Consider $G=(\bigotimes_{2\mathbb{Z}^n} U)$, the operator over $\mathcal{H}$. The induced global evolution is $G$ at odd time steps, and $\sigma G$ at even time steps, where $\sigma$ is a translation by one in all directions (Fig.~\ref{fig:structure}).
\end{Def}

{
\begin{figure}[h!]
\centering
\includegraphics[scale=.71, clip=true, trim=0cm 0cm 0cm 0cm]{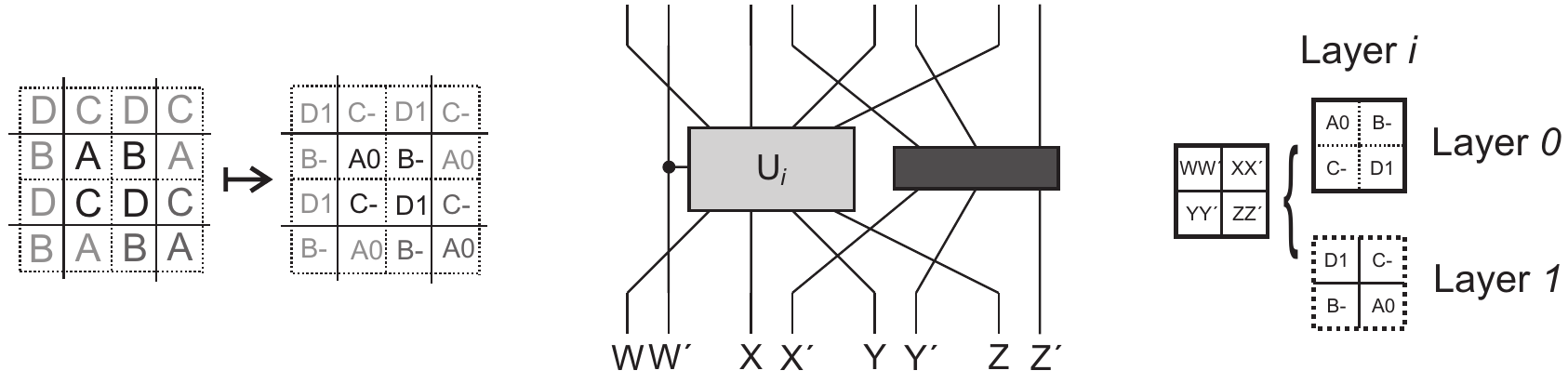}
\caption{PQCA simulating a BQCA. The QCA is decorated with control qubits following a simple encoding procedure (\emph{left}), which allow the scattering unitary $U$ (\emph{centre}) to act as either $U_0$ or $U_1$, according to the layer (\emph{right}). The black box can be any unitary. \label{fig:PQCAbig}}
\end{figure}
}\VS{
\begin{figure}[h!]
\centering
\includegraphics[scale=.7, clip=true, trim=0cm 0cm 0cm 0cm]{img/QCA1new.pdf}
\caption{PQCA simulating a BQCA. The QCA is decorated with control qubits following a simple encoding procedure (\emph{left}), which allow the scattering unitary $U$ (\emph{centre}) to act as either $U_0$ or $U_1$, according to the layer (\emph{right}). The black box can be any unitary. \label{fig:PQCAbig}}
\end{figure}
}
Following previous results  (section \ref{subsec:twolayers}), it is only necessary to show that PQCA can simulate BQCA. Both PQCA and BQCA are two-layer; the only difference is that for BQCA those two layers may be different (\eg compare Figs.~\ref{fig:structure} and \ref{fig:structureBQCA}),
whereas for PQCA there is only a single scattering unitary. So a $U$-defined PQCA, with a $U$ capable of performing $U_0$ and $U_1$ alternatively as controlled by some ancillary suffices. This has been shown for one dimension \cite{ArrighiFI} and is given here for two dimensions in Fig.~\ref{fig:PQCAbig}. It is clear that the construct given here generalises to $n$-dimensions.\\

\begin{Th}[PQCA are universal]\label{th:pqca}
Given any $n$-dimensional QCA $H$, there exists an $n$-dimensional PQCA $G$ which simulates $H$.
\end{Th}

Therefore it can be  concluded that PQCA are the most canonical and general operational description of QCA. More generally, by 
showing here that the various definitions of QCA available  \cite{BrennenWilliams,NagajWocjan,PerezCheung,Karafyllidis,Raussendorf,VanDam,InokuchiMizoguchi,Watrous} are equivalent,
this demonstrates that  a well-axiomatised, concrete, and operational $n$-dimensional QCA is now available.
}


\section{An Intrinsically Universal QCA}\label{sec:nuqca}

\begin{figure}[h]
\centering
\includegraphics[scale=0.75, clip=true, trim=0cm 0cm 0cm 0cm]{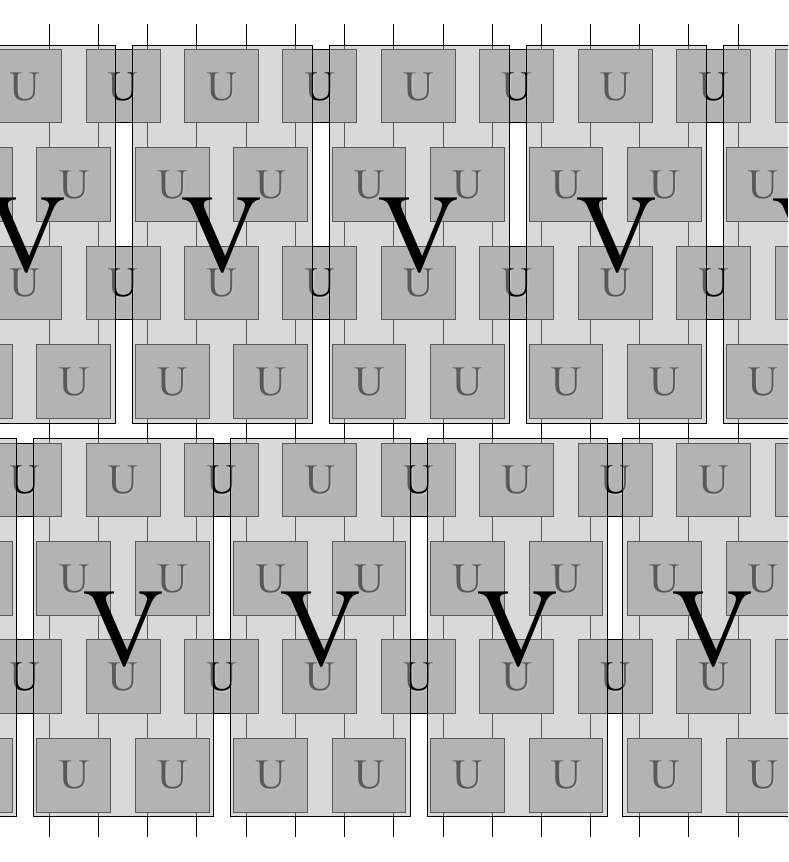}
\caption{Intrinsic simulation of one QCA by another.\label{UsimV}
The QCA defined by $U$ simulates the QCA defined by $V$. In this case two cells of the $U$-defined QCA are required to encode one cell of the $V$-defined QCA, and we need to run the $U$-defined QCA for four time steps to simulate one time step of the $V$-defined QCA. More generally the challenge is to define an initial configuration of the $U$-defined QCA so that it behaves just as the $V$-defined QCA with respect to the encoded initial configuration, after some fixed number of time steps. Such an encoding must hold the configuration of the $V$-defined QCA as well as a way of describing the scattering unitary $V$.}
\end{figure}

\VL{In section \ref{subsecdef} the formal definition of $n$-dimensional PQCA was discussed (Fig.~\ref{fig:structure}), and the formal definition of intrinsic simulation was recalled (Fig.~\ref{UsimV}). }
The aim \VL{now }is to find a particular $U$-defined PQCA which is capable of intrinsically simulating any $V$-defined PQCA, for any $V$. In order to describe such a $U$-defined PQCA in detail, two things must be given: the dimensionality of the cells (including the meaning attached to each of the states they may take), and the way the scattering unitary $U$ acts upon these cells. \VS{\\}\VL{First we discuss the general scheme used to solve this problem, and then we describe the PQCA implementing it.}

\subsection{Circuit Universality versus Intrinsic Universality in Higher Dimensions}
As already discussed, intrinsic universality refers to the ability for one CA to simulate any other CA\VS{,} \VL{in a way which preserves the spatial structure of the simulated CA.
Conversely, computation universality refers to the ability of a CA to simulate any TM, and hence run any algorithm.}\VS{whereas computation universality is about simulating a TM.} Additionally, circuit universality is the ability of one CA to simulate any circuit. \VL{These are \textsc{Nand} gate circuits for classical circuits and classical CA, and \textsc{Toffoli} gate circuits for reversible circuits and reversible CA.} Informally, in a quantum setting, circuit universality is the ability of a PQCA to simulate \VL{any unitary evolution expressed as a}\VS{any finitary} combination of a universal set of quantum gates, such as the standard gate set: \textsc{Cnot, \Phase} (also known as the $\frac{\pi}{8}$ gate), and the \textsc{Hadamard} gate.
\VL{
The relationships between these three concepts of CA universality have been noted previously \cite{DurandRoka}. 
A computation universal CA is also a circuit universal CA, because circuits are finitary computations. 
In addition, an intrinsic universal CA is also a computation universal CA, because it can simulate any CA,
including computation universal CA. Hence intrinsic universality implies computation universality, which implies circuit universality.

In one-dimension this is not an equivalence. Intuitively, computation universality requires more than circuit universality, namely the ability to loop the computation, which is not trivial for CA. Similarly, intrinsic universality requires more than computation universality, such as the ability to simulate multiple communicating TM. In the classical setting there are formal results that distinguish these ideas \cite{OllingerJAC}.}

In $n$-dimensions, it is often assumed in the classical CA literature that circuit universality implies intrinsic universality, and \VL{hence that these are equivalent}\VS{that both are equivalent to computation universality} \cite{OllingerJAC}\VS{, without provision of an explicit construction}. Strictly speaking this is not true. Consider a two-dimensional CA which runs one-dimensional CA in parallel. If the one-dimensional CA is circuit/computation universal, but not computation/intrinsically universal, then this is also true for the two-dimensional CA. Similarly, in the PQCA setting, the two-dimensional constructions in \cite{PerezCheung} and \cite{Raussendorf} are circuit universal but not intrinsically universal.

However, this remains a useful intuition: Indeed, CA admit a block representation, where these blocks are permutations for reversible CA, while for PQCA the blocks are unitary matrices. Thus the evolution of any (reversible/quantum) CA can be expressed as an infinite (reversible/quantum) circuit of (reversible/ quantum) gates repeating across space. If a CA is circuit universal, and if it is possible to wire together different circuit components in different regions of space, then the CA can simulate the block representation of any CA, and hence can simulate any CA in a way which preserves its spatial structure. It is intrinsically universal.\VS{\\}
\VL{
This is the route followed next in constructing the intrinsically universal $n$-dimensional PQCA. First the construction of  the `wires',
which can carry information across different regions of space, is considered. Here these are signals which can be redirected or delayed using barriers, with each signal holding a qubit of information. Secondly, the `circuit-pieces' are constructed, by implementing quantum gates which can be combined. One and two qubit gates are implemented as obstacles to, and interactions of, these signals.}

\subsection{Flattening a PQCA into space}\label{subsec:flat}
\VL{In the classical CA literature it is considered enough to show that the CA implements some wires carrying signals, and some universal gates 
acting upon them, to prove that an $n$-dimensional CA is in fact intrinsically universal. }
\VS{\indent}Any CA can be encoded into a `wire and gates' arrangement following the above argument, but this has never been made explicit in the literature.
This section makes more precise how to flatten any PQCA in space, so that it is simulated by a PQCA which implements quantum
wires and universal quantum gates. Flattening a PQCA means that the infinitely repeating, two-layered circuit is arranged in space so that at the beginning all the signals carrying qubits find themselves in circuit-pieces which implement a scattering unitary of the first layer, and then all synchronously exit and travel to circuit-pieces implementing the scattering unitary of the second layer, etc.
An algorithm for performing this flattening can be provided, however the process will not be described in detail,
for clarity and to follow the classical literature, which largely ignores this process.
 
\begin{figure}
\centering
\VS{\includegraphics[scale=.85, clip=true, trim=0cm 0cm 0cm 0cm]{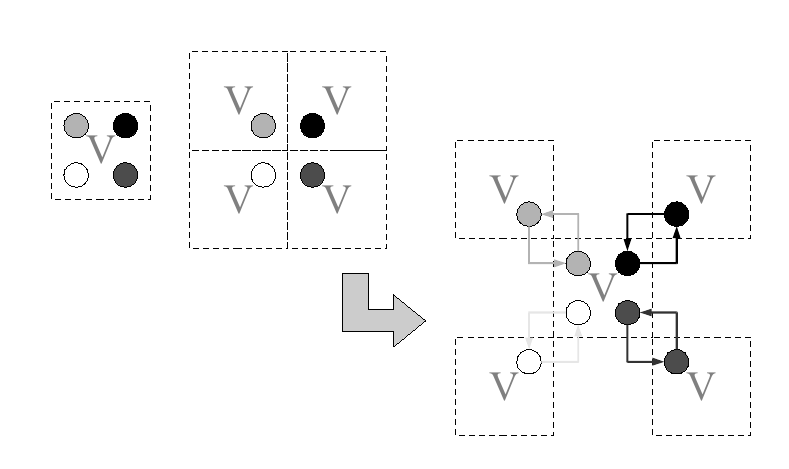}}
\VL{\includegraphics[scale=.9, clip=true, trim=0cm 0cm 0cm 0cm]{img/flattening1and2.pdf}}
\caption{Flattening a PQCA into a simulating PQCA. \emph{Left}: Consider four cells (white, light grey, dark grey, black) of a PQCA having scattering unitary $V$.
The first layer PQCA applies $V$ to these four cells, then the second layer applies $V$ at the four corners. \emph{Right}: We need to flatten this so that the two-layers become non-overlapping. The first layer corresponds to the centre square, and the second layer to the four corner squares. At the beginning the signals (white, light grey, dark grey, black) coding for the simulated cells are in the centre square. \VL{They undergo $V$, and are directed towards the bottom left, top left, bottom right, and top right squares respectively, where they undergo $V$ but paired up with some other signals, etc.} 
\label{fig:flattening12}}
\end{figure}

\begin{figure}
\centering
\VS{\includegraphics[scale=.8, clip=true, trim=0cm 0cm 0cm 0cm]{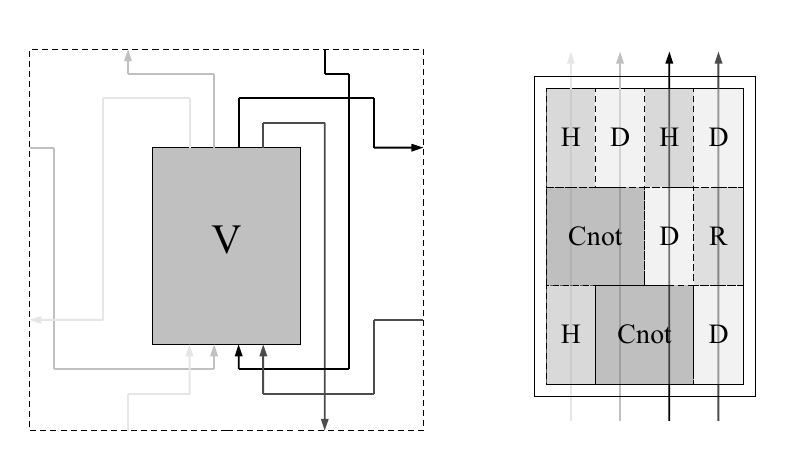}}
\VL{\includegraphics[scale=.9, clip=true, trim=0cm 0cm 0cm 0cm]{img/flattening3and4.pdf}}
\caption{Flattening a PQCA into a simulating PQCA (cont'd). \emph{Left}: Within the central square the incoming signals are bunched together so as to undergo a circuit which implements $V$, and are then dispatched towards the four corners. This diagram does not make explicit a number of signal delays, which may be needed to ensure that they arrive synchronously at the beginning of the circuit implementing $V$. \emph{Right}: Within the central rectangle, the circuit which implements $V$ is itself a combination of smaller circuits for implementing a universal set of quantum gates such as \textsc{Cnot}, \textsc{Hadamard} and the \textsc{\Phase}, together with delays. \VL{These are implemented as explained in sections~\ref{subsec:onequbit} and \ref{subsec:gates}.}\label{fig:flattening34}}
\end{figure}

The flattening process can be expressed in three steps:
Firstly, the $V$-defined PQCA is expanded in space by coding each cell into a hypercube of $2^n$ cells. This allows enough space for the scattering unitary $V$ to be applied on non-overlapping hypercubes of cells,  illustrated in the two-dimensional case in Fig.~\ref{fig:flattening12}.
\noindent Secondly, the hypercubes where $V$ is applied must be connected with wires, as shown in Fig.~\ref{fig:flattening12} $(right)$. Within these hypercubes wiring is required so that incoming signals are bunched together to undergo a circuit implementation of $V$, and are then dispatched appropriately, as shown in Fig.~\ref{fig:flattening34} $(left)$. This requires both time and space expansions, with factors that depend non-trivially (but uninterestingly) upon the size of the circuit implementation of $V$ and the way the wiring and gates work in the simulating PQCA.
\noindent Next, an encoding of the circuit description of the scattering unitary $V$ is implemented in the simulating PQCA upon these incoming bunched wires, as shown in Fig.~\ref{fig:flattening34} $(right)$. This completes the description of the overall scheme according to which a PQCA that is capable of implementing wires and gates is also capable of intrinsically simulating any PQCA, and hence any QCA. A particular PQCA that supports these wires and gates can now be constructed.\VS{\\}

\subsection{Barriers and Signals Carrying Qubits}\label{subsec:onequbit}
Classical CA studies often refer to `signals' without an explicit definition. In this context,
a signal refers to the state of a cell which may move to a neighbouring cell consistently, from one step to another, by the evolution of the CA.
Therefore a signal would appear as a line in the space-time diagram of the CA. These lines need to be implemented as signal redirections. 
A $2$D solution is presented here, but this scheme can easily be extended to higher dimensions. Each cell has four possible basis states: 
empty ($\epsilon$), holding a qubit signal ($0$ or $1$), or a barrier ($\blacksquare$). The scattering unitary $U$ of the universal PQCA acts on $2\times 2$ cell neighbourhoods.

Signals encode qubits which can travel diagonally across the 2D space (NE, SE, SW, or NW).
Barriers do not move, while signals move in the obvious way if unobstructed, as there is only one choice for any signal in any square of four cells. 
Hence the basic movements of signals are given by the following four rules:
$$\cells{}{}{$s$}{} \mapsto \cells{}{$s$}{}{}, \qquad \cells{$s$}{}{}{} \mapsto \cells{}{}{}{$s$},$$
$$\cells{}{$s$}{}{} \mapsto \cells{}{}{$s$}{}, \qquad \cells{}{}{}{$s$} \mapsto \cells{$s$}{}{}{}.$$
where $s\in \{0,1\}$ denotes a signal, and blank cells are empty. \\
The four rules above should be interpreted in as a case-by-case definition of the scattering unitary $U$, 
\ie they show that 
$U\cells{}{}{$s$}{}=\cells{}{$s$}{}{}$.
As each rule can be obtained as a rotation of any other, by stating that the $U$-defined PQCA is isotropic the first rule above suffices. This convention will be used throughout.

\begin{figure}
\centering
\VS{\includegraphics[scale=.55, clip=true]{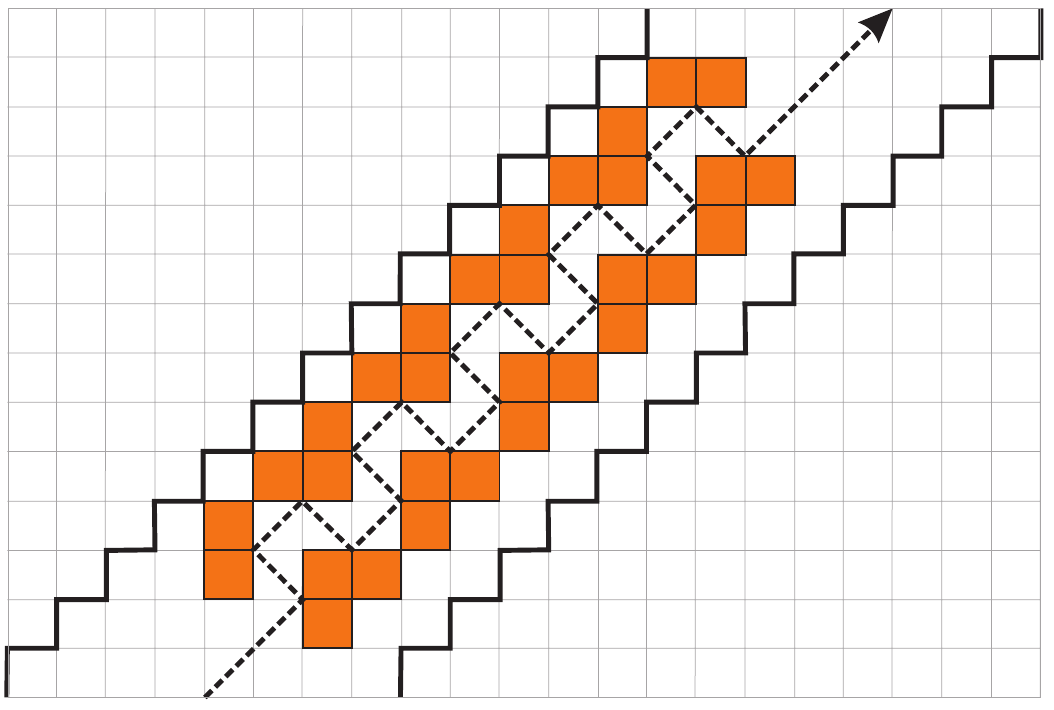}}
\VL{\includegraphics[scale=.60, clip=true]{img/delayCirc.pdf}}
\caption{The `identity circuit' tile, an $8\times 14$ tile taking 24 time-steps, made by repeatedly bouncing the signal
from walls to slow its movement through the tile. The dotted line gives the signal trajectory, with the arrow showing the 
exit point and direction of signal propagation. The bold lines show the tile boundary.}
\label{fig:delays}
\end{figure}

\begin{figure}
\centering
\VS{\includegraphics[scale=.55, clip=true]{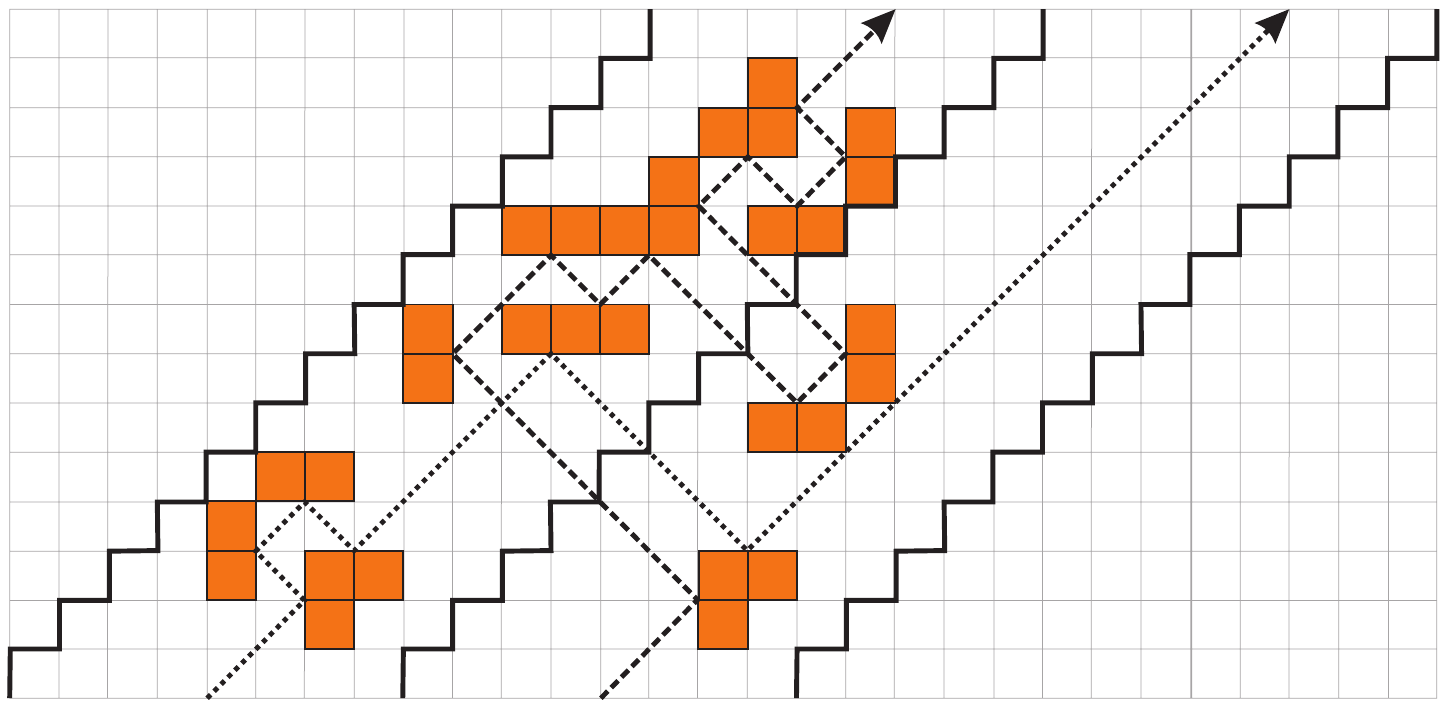}}
\VL{\includegraphics[scale=.60, clip=true]{img/swapCirc.pdf}}
\caption{The `swap circuit' tile, a $16\times 14$ tile, where both input signals are permuted and exit synchronously after 24 time-steps.
As the first signal (\emph{bottom left}) is initially delayed, there is no interaction.}
\label{fig:swap}
\end{figure}

The ability to redirect signals is achieved by `bouncing' them off walls constructed from
two barriers arranged either horizontally or vertically:
$$
\cells{\barrier}{$s$}{\barrier}{} \mapsto \cells{\barrier}{}{\barrier}{$s$}.
$$
where $s$ again denotes the signal and the shaded cells denote the barriers which causes the signal 
to change direction.
If there is only one barrier present in the four cell square being operated on then the signal simply propagates as normal and is not deflected:
$$\cells{\barrier}{}{$s$}{} \mapsto \cells{\barrier}{$s$}{}{}.$$
Using only these basic rules of signal propagation and signal reflection from barrier walls, signal delay (Fig.  \ref{fig:delays}) and signal swapping (Fig.~\ref{fig:swap}) 
tiles can be constructed. All of the rules presented so far are permutations of some of the base elements of the vector space generated by
$$\Set{\cells{$w$}{$x$}{$y$}{$z$}}_{w,x,y,z \in \{\epsilon,0,1,\blacksquare\}}$$
therefore $U$ is indeed unitary on the subspace upon which its action has so far been described. 

\subsection{Gates}\label{subsec:gates}
\begin{figure}
\centering
\VS{\vspace{-0.5mm}\includegraphics[scale=.55, clip=true]{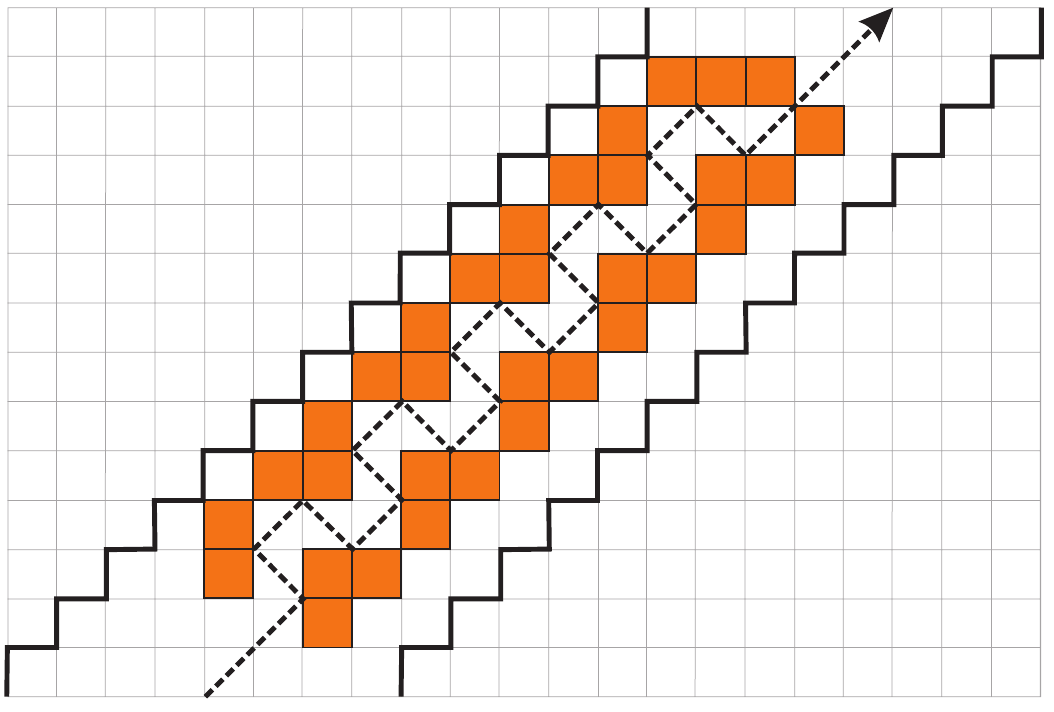}}
\VL{\includegraphics[scale=.60, clip=true]{img/hadCirc.pdf}}
\caption{The `Hadamard gate' tile applies the Hadamard operation to the input signal. It is a modification
of the identity circuit tile, with a diagonal (semitransparent) barrier added at the end which performs the Hadamard operation.}
\label{fig:hadamard}
\end{figure}

\begin{figure}
\centering
\VS{\vspace{-0.5mm}\includegraphics[scale=.55, clip=true]{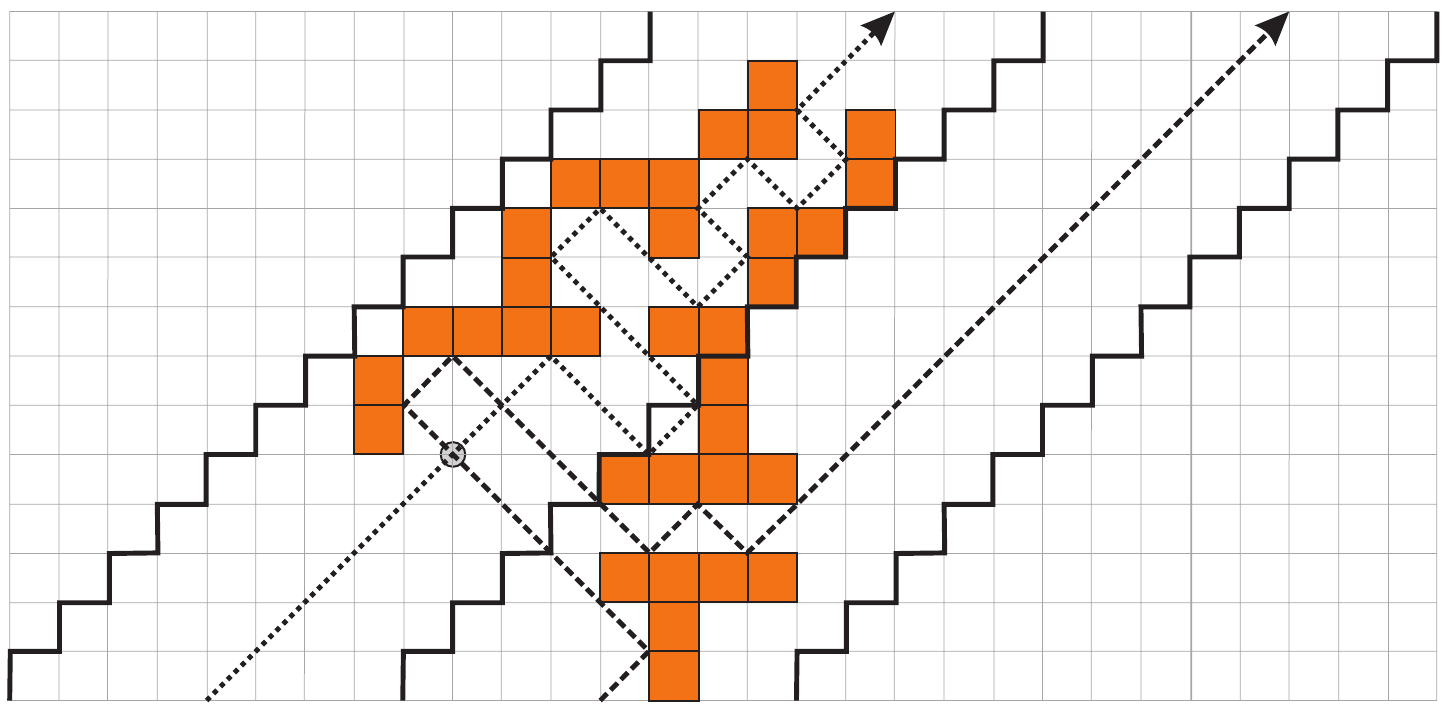}}
\VL{\includegraphics[scale=.60, clip=true]{img/cPhaseCirc.pdf}}
\caption{The `\cPhase{} gate' tile\VS{, with a signal interaction at the highlighted cell.}\VL{ applies the controlled-\Phase{} operation to the two input qubits, by causing the signals to interact at the highlighted point (grey circle). The qubits are then synchronised so that they exit at the same time along their original paths. No swapping takes place.}}
\label{fig:cphase}
\end{figure}

\begin{figure}
\centering
\VS{\vspace{-0.5mm}\includegraphics[scale=.55, clip=true]{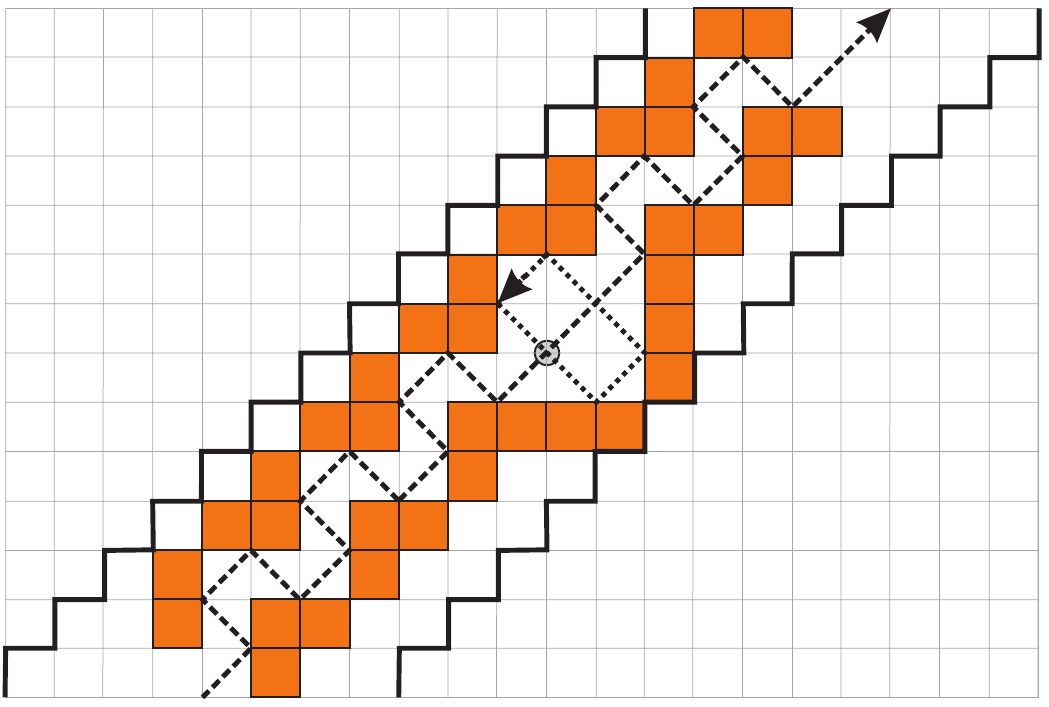}}
\VL{\includegraphics[scale=.60, clip=true]{img/phaseCirc.pdf}}
\caption{The `\Phase{} gate' tile. This tile makes use
of a signal, set to $\ket{1}$,  which loops inside the grid every six time-steps, ensuring that it will interact with the signal that enters the
tile, and causing it to act as the control qubit to a \cPhase{} operation. It therefore acts as a phase rotation on the input qubit, which passes directly through. \VL{After 24 time-steps the auxiliary control signal has returned to its origin, unchanged, hence the tile can be reused.}}
\label{fig:phase}
\end{figure}

To allow a universal set of gates to be implemented by the PQCA,  certain combinations
of signals and barriers can be assigned special importance.
The Hadamard operation on a single qubit-carrying signal can be implemented by interpreting a
signal passing through a diagonally oriented wall, analogous to a semitransparent barrier in physics. This has the action
defined by the following rule:
$$\cells{\barrier}{}{0}{\barrier} \mapsto\frac{1}{\sqrt{2}}\cells{\barrier}{0}{}{\barrier} + \frac{1}{\sqrt{2}}\cells{\barrier}{1}{}{\barrier}$$
$$\cells{\barrier}{}{1}{\barrier} \mapsto \frac{1}{\sqrt{2}}\cells{\barrier}{0}{}{\barrier} - \frac{1}{\sqrt{2}}\cells{\barrier}{1}{}{\barrier} $$
This implements the Hadamard operation, creating a superposition of configurations with appropriate phases. Using this construction
a Hadamard tile can be constructed (Fig.~\ref{fig:hadamard}) by simply adding a semitransparent barrier to the end of the previously
defined delay (identity) tile (Fig.~\ref{fig:delays}).
A way of encoding two qubit gates in this system is to consider that two signals which
cross paths interact with one another. The controlled-\Phase{ }
operation can be implemented by considering signals that cross each other as interacting only if they are both $1$, in which case a global phase
of $e^{\frac{i\pi}{4}}$ is applied. Otherwise the signals continue as normal. This behaviour is defined by the following rule:
$$\cells{1}{}{1}{} \mapsto e^{\frac{i\pi}{4}}\cells{}{1}{}{1}, \qquad \cells{$x$}{}{$y$}{} \mapsto \cells{}{$y$}{}{$x$} otherwise$$
where $x,y \in \{0,1\}$. This signal interaction which induces a global phase change allows the definition of both a two signal
controlled-\Phase{} tile (Fig.~\ref{fig:cphase}) and a single signal \Phase{} operation tile (Fig.~\ref{fig:phase}).
These rules are simply a permutation and phase change of base elements of the form: 
$$\Set{\cells{$x$}{}{$y$}{}}_{x,y \in \{0,1\}}$$
(and their rotations), therefore $U$ is a unitary operation on the subspace upon which its action has so far been described.
Wherever $U$ has not yet been defined, it is the identity. Hence $U$ is unitary.\VS{\\} 

\subsection{Circuits: Combining Gates}\label{subsec:circuits}
A signal is given an $8 \times 14$ tile ($16 \times 14$ for two signal operations)
in which the action is encoded. The signals enter each tile at the fifth cell from the left, and propagate diagonally NE.
Each time step finds the tile shifted one cell to the right to match this diagonal movement, giving a diagonal tile.
The signal exits the tile $14$ cells North and East of where it entered. This allows
these tiles to be composed in parallel and sequentially with the only other requirement being that the signal exits at the appropriate point, \ie
the fifth cell along the tile, after $24$ time-steps. This ensures that all signals are synchronised as in Fig.~\ref{fig:flattening34} (\emph{right}),
allowing larger circuits to be built from these elementary tiles by simply plugging them together. Non-contiguous gates can also be wired together
using appropriate wall constructions to redirect and delay signals so that they are correctly synchronised.

The implemented set of quantum gates, the identity, Hadamard, swap, \Phase{} and \cPhase{}, gives a universal set. Indeed the standard set
of \textsc{cNot}, \textsc{H}, \textsc{\Phase} can be recovered as follows:
$$\textsc{cNot}\ket{\psi}=(\mathbb{I}\otimes H)(\textsc{cR(${\pi}\slash{4}$)})^4(\mathbb{I}\otimes H)\ket{\psi}$$
where $\textsc{cR($\frac{\pi}{4}$})^4$ denotes four applications of the \cPhase{} gate, giving the controlled-\textsc{Phase} operation.

\section{Conclusion}\label{sec:discussion}

{\em Summary.} This paper defines and promotes $n$-dimensional intrinsic universality as a useful concept, and also proves two concrete results: that PQCA are intrinsically universal, and that there exists a universal instance of them. There are several consequences, summarised here:
\begin{itemize}
\item The construction demonstrates that all the non-axiomatic definitions of QCA \cite{BrennenWilliams,NagajWocjan,PerezCheung,Karafyllidis,Raussendorf,VanDam,WatrousFOCS,InokuchiMizoguchi} are equivalent to one another and to the axiomatic definition, \ie they all simulate each other. This therefore demonstrates that the concept of $n$-dimensional QCA is well-axiomatised, concrete, and operational. 
\item The QCA model is simplified, \ie without loss of generality QCA can be assumed to be a PQCA (see Def.~\ref{def:pqca}), or even a specific instance of PQCA (section~\ref{sec:nuqca}).
\end{itemize}
This paper also presents a simple PQCA which is capable of simulating all other PQCA, preserving the topology of the simulated PQCA. This means that the initial configuration and the forward evolution of any PQCA can be encoded within the initial configuration of this PQCA, with each simulated cell encoded as a group of adjacent cells in the PQCA, \ie intrinsic simulation. The construction in section \ref{sec:nuqca} is given in two-dimensions, which can be seen to generalise to $n>1$ dimensions. This second result can therefore be stated as follows: 

\emph{There exists an $n$-dimensional $U$-defined PQCA, $G$, which is an intrinsically universal PQCA. Let $H$ be a $n$-dimensional $V$-defined PQCA such that $V$ can be expressed as a quantum circuit $C$ made of gates from the set $\textsc{Hadamard}$, $\textsc{Cnot}$, and $\textsc{\Phase}$. Then $G$ is able to intrinsically simulate $H$.}

Any finite-dimensional unitary $V$ can always be approximated by a circuit $C(V)$ with an arbitrary small error $\varepsilon=\max_{\ket{\psi}}||V\ket{\psi}-C\ket{\psi}||$. Assuming instead that $G$ simulates the $C(V)$-defined PQCA, for a region of $s$ cells over a period $t$, the error with respect to the $V$-defined PQCA will be bounded by $st\varepsilon$. This is due to the general statement that errors in quantum circuits increase, at most, proportionally with time and space \cite{NielsenChuang}.
\VS{\\} 

{\em Discussion.} QC research has so far focused on applications for more secure and efficient computing, with theoretical physics supporting this work in theoretical computer science. The results of this interdisciplinary exchange led to the assumptions underlying computer science to be revisited,
with information theory and complexity theory, for example, being reconsidered and redeveloped.
However, information theory also plays a crucial role in the foundations of theoretical physics\VS{ (\eg deepening our understanding of entanglement \cite{DurVidalCirac} and decoherence \cite{PazZurek})}. 
These developments are also of interest in theoretical physics studies where physical aspects such as particles and matter are considered. Computer science studies can help to consider these as abstract mathematical quantities only. 
Universality, among the many  concepts in computer science, is a simplifying methodology in this respect. For example, if the problem being studied crucially involves some idea of interaction, universality makes it possible to cast it in terms of information exchanges \emph{together} with some universal information processing.
This paper presents an attempt to export universality as a tool for application in theoretical physics; a small step towards the goal of finding and understanding a  \emph{universal physical phenomenon}, within some simplified mechanics.
Similar to the importance of the idea of the spatial arrangement of interactions in physics, intrinsic universality has broader applicability than computation universality and must be preferred. In short, if only one physical phenomenon is considered, it should be an intrinsically universal physical phenomenon, as it could be used to simulate all others.

Moreover, the intrinsic universality of PQCA developed here could be given a physical interpretation. QCA, as seen through their axiomatic definition (Def.~\ref{def:qca}), are synonymous with discrete-time, discrete-space quantum mechanics (together with some extra assumptions such as translation-invariance and finite-density of information). Stating that discrete-time, discrete-space quantum mechanical evolutions can, without loss of generality, be assumed to be of the form illustrated in Fig.~\ref{fig:structure}, amounts to the statement that `scattering phenomena are universal physical phenomena'. In this sense, the result leads to an understanding of the links between the axiomatic, top-down principles approach to theoretical physics, and the more bottom-up study of the scattering of particles.

\VS{\vspace{-4mm}}\section*{Acknowledgements}
\VS{\vspace{-2mm}}The authors would like to thank  J\'er\^ome Durand-Lose, Jarkko Kari, Jacques Mazoyer, Kenichi Morita, Nicolas Ollinger, Guillaume Theyssier and Philippe Jorrand.
\bibliography{../../../Bibliography/biblio}

\begin{thebibliography}{10}
\expandafter\ifx\csname url\endcsname\relax
  \def\url#1{\texttt{#1}}\fi
\expandafter\ifx\csname urlprefix\endcsname\relax\def\urlprefix{URL }\fi
\expandafter\ifx\csname href\endcsname\relax
  \def\href#1#2{#2} \def\path#1{#1}\fi

\bibitem{Neumann}
J.~von Neumann, {Theory of Self-Reproducing Automata}, University of Illinois
  Press, Champaign, IL, USA, 1966.

\bibitem{NeumannQT}
J.~von Neumann, {Mathematical foundations of quantum mechanics}, Princeton
  University Press, 1955.

\bibitem{FeynmanQC}
R.~P. Feynman, {Simulating physics with computers}, International Journal of
  Theoretical Physics 21~(6) (1982) 467--488.

\bibitem{BrennenWilliams}
G.~K. Brennen, J.~E. Williams, {Entanglement dynamics in one-dimensional
  quantum cellular automata}, Phys. Rev. A 68~(4) (2003) 042311.
\newblock \href {http://dx.doi.org/10.1103/PhysRevA.68.042311}
  {\path{doi:10.1103/PhysRevA.68.042311}}.

\bibitem{LloydQCA}
S.~Lloyd, {A potentially realizable quantum computer}, Science 261~(5128)
  (1993) 1569--1571.

\bibitem{NagajWocjan}
D.~Nagaj, P.~Wocjan, {Hamiltonian Quantum Cellular Automata in 1D}, ArXiv
  preprint: arXiv:0802.0886 (2008).

\bibitem{Twamley}
J.~Twamley, {Quantum cellular automata quantum computing with endohedral
  fullerenes}, Phys. Rev. A 67~(5) (2003) 52318--52500.

\bibitem{VollbrechtCirac}
K.~G.~H. Vollbrecht, J.~I. Cirac, {Reversible universal quantum computation
  within translation-invariant systems}, New J. Phys Rev A 73 (2004) 012324.

\bibitem{Bialynicki-Birula}
I.~Bialynicki-Birula, {Weyl, Dirac, and Maxwell equations on a lattice as
  unitary cellular automata}, Physical Review D 49~(12) (1994) 6920--6927.

\bibitem{BoghosianTaylor2}
B.~M. Boghosian, W.~Taylor, {Quantum lattice-gas model for the many-particle
  Schr{\"o}dinger equation in d dimensions}, Phys. Rev. E 57~(1) (1998) 54--66.

\bibitem{Eakins}
J.~Eakins, {Quantum cellular automata, the EPR paradox and the Stages
  paradigm}, in: Proceedings of NATO ARW, The Nature of Time: Geometry, Physics
  and Perception, 2003, p. 323.

\bibitem{LoveBoghosian}
P.~Love, B.~Boghosian, {From Dirac to Diffusion: decoherence in Quantum Lattice
  gases}, Quantum Information Processing 4 (2005) 335--354.

\bibitem{MeyerQLGI}
D.~A. Meyer, {From quantum cellular automata to quantum lattice gases}, J.
  Stat. Phys 85 (1996) 551--574.

\bibitem{MargolusPhysics}
N.~Margolus, {Physics-like models of computation}, Physica D: Nonlinear
  Phenomena 10~(1-2).

\bibitem{MargolusQCA}
N.~Margolus, {Parallel quantum computation}, in: Complexity, Entropy, and the
  Physics of Information: The Proceedings of the 1988 Workshop on Complexity,
  Entropy, and the Physics of Information Held May-June, 1989, in Santa Fe, New
  Mexico, Perseus Books, 1990, p. 273.

\bibitem{Codd}
E.~G. Codd, {Cellular Automata}, Academic Press, New York, 1978.

\bibitem{MazoyerFiring}
J.~Mazoyer, {A Six-State Minimal Time Solution to the Firing Squad
  Synchronization Problem}, Theoretical Computer Science 50 (1987) 183--238.

\bibitem{MooreFiring}
E.~F. Moore, {The firing squad synchronization problem}, in: Sequential
  Machines, Selected Papers, Addison-Wesley, 1964, pp. 213--214.

\bibitem{ToffoliMargolusModelling}
T.~Toffoli, N.~Margolus, {Cellular Automata Machine --- A new Environment for
  Modelling}, MIT Press, Cambridge MA, 1987.

\bibitem{LloydQG}
S.~Lloyd, {A theory of quantum gravity based on quantum computation}, ArXiv
  preprint: quant-ph/0501135 (2005).

\bibitem{SchumacherWerner}
B.~Schumacher, R.~Werner, {Reversible quantum cellular automata.}, ArXiv
  pre-print quant-ph/0405174 (2004).

\bibitem{ArrighiLATA}
P.~Arrighi, V.~Nesme, R.~F. Werner, {Quantum cellular automata over finite,
  unbounded configurations}, in: Proceedings of MFCS, Lecture Notes in Computer
  Science, Vol. 5196, Springer, 2008, pp. 64--75.

\bibitem{ArrighiUCAUSAL}
P.~Arrighi, V.~Nesme, R.~Werner, {Unitarity plus causality implies
  localizability}, {QIP 2010 and Journal of Computer and System Sciences, ArXiv
  preprint: arXiv:0711.3975}.

\bibitem{PerezCheung}
C.~P{\'e}rez-Delgado, D.~Cheung, {Local unreversible cellular automaton
  ableitary quantum cellular automata}, Physical Review A 76~(3) (2007) 32320.

\bibitem{Karafyllidis}
I.~G. Karafyllidis, {Definition and evolution of quantum cellular automata with
  two qubits per cell}, Journal reference: Phys. Rev. A 70 (2004) 044301.

\bibitem{Raussendorf}
R.~Raussendorf, {Quantum cellular automaton for universal quantum computation},
  Physical Review A 72~(2) (2005) 22301.

\bibitem{VanDam}
W.~Van~Dam, {Quantum cellular automata}, Masters thesis, University of
  Nijmegen, The Netherlands (1996).

\bibitem{WatrousFOCS}
J.~Watrous, On one-dimensional quantum cellular automata, Foundations of
  Computer Science, Annual IEEE Symposium on 528537 (1995) 528--537.
\newblock \href
  {http://dx.doi.org/http://doi.ieeecomputersociety.org/10.1109/SFCS.1995.4925%
83} {\path{doi:http://doi.ieeecomputersociety.org/10.1109/SFCS.1995.492583}}.

\bibitem{InokuchiMizoguchi}
S.~Inokuchi, Y.~Mizoguchi, {Generalized partitioned quantum cellular automata
  and quantization of classical CA}, International Journal of Unconventional
  Computing, ArXiv preprint: quant-ph/0312102 1 (2005) 149--160.

\bibitem{winningways}
E.~Berlekamp, J.~Conway, R.~Guy, {Winning ways for your mathematical plays}, AK
  Peters, Ltd., 2003.

\bibitem{AlbertCulik}
J.~Albert, K.~Culik, {A simple universal cellular automaton and its one-way and
  totalistic version}, Complex Systems 1 (1987) 1--16.

\bibitem{Banks}
E.~R. Banks, {Universality in cellular automata}, in: SWAT '70: Proceedings of
  the 11th Annual Symposium on Switching and Automata Theory (SWAT 1970), IEEE
  Computer Society, Washington, DC, USA, 1970, pp. 194--215.
\newblock \href {http://dx.doi.org/http://dx.doi.org/10.1109/SWAT.1970.27}
  {\path{doi:http://dx.doi.org/10.1109/SWAT.1970.27}}.

\bibitem{DurandRoka}
B.~Durand, Z.~Roka, {The Game of Life: universality revisited Research Report
  98-01}, Tech. rep., Ecole Normale Suprieure de Lyon (1998).

\bibitem{MazoyerRapaport}
J.~Mazoyer, I.~Rapaport, {Inducing an order on cellular automata by a grouping
  operation}, in: Proceedings of STACS'98, in Lecture Notes in Computer
  Science, Vol. 1373, Springer, 1998, pp. 116--127.
\newblock \href {http://dx.doi.org/10.1007/BFb0028542}
  {\path{doi:10.1007/BFb0028542}}.

\bibitem{OllingerJAC}
N.~Ollinger, {Universalities in cellular automata a (short) survey.}, in:
  B.~Durand (Ed.), First Symposium on Cellular Automata ``Journ{\'e}es
  Automates Cellulaires'' (JAC 2008), Uz{\`e}s, France, April 21-25, 2008.
  Proceedings, MCCME Publishing House, Moscow, 2008, pp. 102--118.

\bibitem{Theyssier}
G.~Theyssier, {Captive cellular automata}, in: Proceedings of MFCS 2004, in
  Lecture Notes in Computer Science, Vol. 3153, Springer, 2004, pp. 427--438.

\bibitem{ArrighiFI}
P.~Arrighi, R.~Fargetton, Z.~Wang, {Intrinsically universal one-dimensional
  quantum cellular automata in two flavours}, Fundamenta Informaticae 21 (2009)
  1001--1035.

\bibitem{MoritaCompUniv1D}
K.~Morita, M.~Harao, {Computation universality of one-dimensional reversible
  (injective) cellular automata}, IEICE Trans. Inf. \& Syst., E 72 (1989)
  758--762.

\bibitem{MoritaCompUniv2D}
K.~Morita, S.~Ueno, {Computation-universal models of two-dimensional 16-state
  reversible cellular automata}, IEICE Trans. Inf. \& Syst., E 75 (1992)
  141--147.

\bibitem{MoritaIntrinsicUniv1D}
K.~Morita, {Reversible simulation of one-dimensional irreversible cellular
  automata}, Theoretical Computer Science 148~(1) (1995) 157--163.

\bibitem{Durand-LoseLATIN}
J.~O. Durand-Lose, {Reversible cellular automaton able to simulate any other
  reversible one using partitioning automata}, in: In {LATIN'95}: Theoretical
  Informatics, number 911 in Lecture Notes in Computer Science, Springer, 1995,
  pp. 230--244.

\bibitem{Durand-LoseIntrinsic1D}
J.~O. Durand-Lose, {Intrinsic universality of a 1-dimensional reversible
  cellular automaton}, in: Proceedings of STACS 97, Lecture Notes in Computer
  Science, Springer, 1997, p. 439.

\bibitem{ToffoliConstruction}
T.~Toffoli, {Computation and construction universality of reversible cellular
  automata}, J. of Computer and System Sciences 15~(2).

\bibitem{KariCircuit}
J.~Kari, {On the circuit depth of structurally reversible cellular automata},
  Fundamenta Informaticae 38~(1-2) (1999) 93--107.

\bibitem{ShepherdFranz}
D.~J. Shepherd, T.~Franz, R.~F. Werner, {A universally programmable quantum
  cellular automata}, Phys. Rev. Lett. 97~(020502).

\bibitem{Durand-LoseEnc}
J.~O. Durand-Lose, {Universality of Cellular Automata}, in: Encyclopedia of
  Complexity and System Science, Springer, 2008, p.~22.

\bibitem{ArrighiPQCA}
P.~Arrighi, J.~Grattage, {Partitioned quantum cellular automata are
  intrinsically universal}, accepted for publication, Post-proceedings of the
  Physics and Computation workshop (2009).

\bibitem{ArrighiSimple}
P.~Arrighi, J.~Grattage, {A Simple $n$-Dimensional Intrinsically Universal
  Quantum Cellular Automaton}, Language and Automata Theory and Applications,
  Lecture Notes in Computer Science 6031 (2010) 70--81.

\bibitem{Hedlund}
G.~A. Hedlund, {Endomorphisms and automorphisms of the shift dynamical system},
  Math. Systems Theory 3 (1969) 320--375.

\bibitem{DurVidalCirac}
W.~D{\"u}r, G.~Vidal, J.~I. Cirac, {Three qubits can be entangled in two
  inequivalent ways}, Phys. Rev. A 62 (2000) 062314.

\bibitem{PazZurek}
J.~P. Paz, W.~H. Zurek, {Environment-induced decoherence and the transition
  from quantum to classical}, Lecture Notes in Physics (2002) 77--140.

\bibitem{Watrous}
J.~Watrous, {On one-dimensional quantum cellular automata}, Complex Systems
  5~(1) (1991) 19--–30.

\bibitem{NielsenChuang}
M.~A. Nielsen, I.~L. Chuang, {Quantum Computation and Quantum Information},
  {Cambridge University Press}, 2000.

\end{thebibliography}
\bibliographystyle{elsarticle-num}
\end{document}